\documentclass[12pt]{article}
\usepackage{amssymb,amsthm,amsmath}
\usepackage{graphicx}

\renewcommand{\baselinestretch}{1.1}
\theoremstyle{plain}
\newtheorem{proposition}{Proposition}[section]
\newtheorem{theorem}[proposition]{Theorem}
\newtheorem{lemma}[proposition]{Lemma}
\newtheorem{corollary}[proposition]{Corollary}
\newtheorem{conjecture}{Conjecture}

\theoremstyle{definition}
\newtheorem{definition}[proposition]{Definition}
\newtheorem{example}[proposition]{Example}
\newtheorem{remark}[proposition]{Remark}
\newtheorem{claim}[proposition]{Claim}

\def\IR{\mathbb{R}}
\def\IC{\mathbb{C}}
\def\IZ{\mathbb{Z}}
\def\IP{\mathbb{P}}
\def\cale{\mathcal{E}}
\def\calf{\mathcal{F}}
\def\calh{\mathcal{H}}
\def\cala{\mathcal{A}}
\def\calb{\mathcal{B}}
\def\calo{\mathcal{O}}
\def\caln{\mathcal{N}}
\def\bX{\mathbf X}

\def\Ext{\mbox{Ext}}
\def\Hom{\mbox{Hom}}
\def\RHom{\mbox{RHom}}
\def\Tr{\mbox{Tr}}
\def\ch{\mbox{ch}}
\def\deg{\mbox{deg}}
\def\Td{\mbox{Td}}
\def\id{\mbox{id}}
\def\sign{\mbox{sign}}
\def\be{\begin{equation}}
\def\ee{\end{equation}}
\def\iff{\Leftrightarrow}

\begin{document}
\begin{titlepage}

\begin{flushright}
hep-th/0405118\\
NSF-KITP-04-53
\end{flushright}
\vfil

\begin{center}
{\huge Seiberg Duality is an Exceptional Mutation }\\
\vspace{3mm}
%{\huge }

\end{center}

\vfil
\begin{center}
{\large Christopher P. Herzog}\\
\vspace{1mm}
KITP, UCSB, \\ 
Santa Barbara, CA  93106, U.S.A.\\
{\tt herzog@kitp.ucsb.edu}\\
\vspace{3mm}
\end{center}

\vfil

\begin{center}
{\large Abstract}
\end{center}

\noindent
The low energy
gauge theory living on D-branes probing a del Pezzo
singularity of a non-compact Calabi-Yau manifold is not unique.
In fact there is a large equivalence class of such gauge theories
related by Seiberg duality.
As a step toward characterizing this class, 
we show that Seiberg duality can be defined consistently
as an admissible mutation of a strongly exceptional collection
of coherent sheaves.  
\vfil
\begin{flushleft}
May 2004
\end{flushleft}
\vfil
\end{titlepage}
\newpage
\renewcommand{\baselinestretch}{1.1}  %looks better

%%%%%%%%%%%%%%%%%%%%%%%%%%%%%%%%%%%%%%%%%%%%%
%% include the next line for double spacing %%
%%%%%%%%%%%%%%%%%%%%%%%%%%%%%%%%%%%%%%%%%%%%%%
%\renewcommand{\baselinestretch}{2}

\section{Introduction}

We study the action of 
Seiberg duality on 
strongly
coupled gauge theories with gravity duals.  
Our examples are constructed from
D-branes probing del Pezzo
singularities of a non-compact Calabi-Yau
manifold.

These del Pezzo constructions are interesting examples
of gauge/gravity duality because they exhibit several
new features not observed in simpler models.  For example,
\cite{HananyWalcher, Hananywall, chaos} gave evidence that
the gauge theories for these del Pezzo models often have
duality walls.  In other words, the gauge theories cannot
not be defined at energies higher than the scale of the wall.
In \cite{chaos}, the authors argue that the renormalization
group flows for these models 
not only have walls but also can be
chaotic.

The motivation for this paper comes partially from a desire
to better understand the duality walls and chaotic renormalization
group flows of these del Pezzo theories.  In order to follow the renormalization
group flow, the authors of \cite{Hananywall, chaos}  relied on ideas
of Klebanov and Strassler (KS) 
\cite{KlebanovStrassler}.  KS studied D-branes probing a conifold
singularity and discovered that the renormalization group flows
could be continued past strong coupling by looking at the Seiberg dual.
One crucial difference between the KS paper and the del Pezzo examples is
that while the Seiberg dual of the conifold gauge theory is essentially
the same gauge theory, Seiberg duality acting on a del Pezzo gauge theory
produces, typically, an infinite tree of new theories.
To understand duality walls and chaotic renormalization group flows,
we first need a better understanding of the different possible Seiberg dual theories.

A second motivation for the present paper is a desire to clear up some puzzles
in the immense literature on Seiberg duality for del Pezzos.  
In addition to Seiberg's original gauge theory analysis \cite{Seibergduality}
(see \cite{IntSei} for a review), here are four conjectured alternate ways
of understanding the equivalence:
\begin{enumerate}
\item Toric Duality: Using toric geometry to derive del Pezzo gauge theories,
the authors of \cite{BP, HananySD} noticed that
a phenomenon called toric duality could example by example be understood
as Seiberg duality.  However, not all del Pezzos are toric.
\item Picard-Lefschetz Monodromy: 
The authors of \cite{HananySD, CFIKV, oova} pointed out that certain
but not all Picard-Lefschetz monodromies in the mirror geometry were
Seiberg dualities.  
\item Mutation: Using exceptional collections of sheaves
to generate the gauge theory, 
\cite{CFIKV} and \cite{Wijnholt}
noticed that certain combinations of mutations (or braiding operations
on the collections) were Seiberg dualities.
The combinations that were not mutations were labelled 
partial Seiberg dualities.  On an example by example basis,
partial Seiberg dualities tend to produce pathological gauge theories
\cite{HananyPL, Herzog}.
\item Tilting Equivalences: The low energy 
gauge theory on these 
D-brane configurations can be described by a quiver
with relations.  There is a derived category which
can be constructed from the quiver.
Berenstein and Douglas \cite{BD} conjecture that Seiberg
duality can be understood as a tilting equivalence
of the quiver derived category.  This idea was later
developed by \cite{Braun, indians}.
\end{enumerate}

This brief list makes several problems clear.
In all this early work, although some Seiberg dualities had 
alternate interpretations, it was never clear that all Seiberg dualities
could be understood using some other framework.  It was also not clear
what role if any partial Seiberg dualities should play.  
Here are two important questions:
How 
general is the equivalence between Seiberg duality
and any item on this list?
How precisely are the various items on this list related to each other?

In this paper, we show that mutation is a 
consistent alternate way of thinking about Seiberg duality
for all our del Pezzo examples.  Our results 
suggest a close relationship between mutation and tilting equivalences.

As a step toward classifying all possible Seiberg dual theories for del Pezzos,
we find evidence that all these theories are well split.  A precise definition
begins section 5; here we give a simple but intriguing consequence 
for the superpotential.  Well split means that there
is a cyclic ordering of the $SU(N)$ gauge groups which any possible superpotential
term respects.

%From a physics standpoint, 
%we gain a clearer appreciation of the kinds of
%gauge theory structures that arise from
%placing D-branes at del Pezzo singularities.
%For example, our results suggest such gauge theories
%should be well split \cite{Herzog}.
%From a mathematics point of view, we learn
%about exceptional collections
%of sheaves on del Pezzo surfaces \cite{Rudakov}.  In particular,
%Seiberg duality can be understood as
%an admissible mutation in the language of Bondal \cite{Bondal}.

\subsection{Statement of Results}

Our strongly coupled gauge theories are engineered 
from a ten dimensional geometry $\IR^{3,1} \times \bX$
where $\IR^{3,1}$ is four dimensional Minkowski space
and $\bX$ is a three complex dimensional Calabi-Yau.
We consider a severely restricted class of $\bX$.  
Let $\calb$
be a del Pezzo surface, i.e.
$\IP^2$, $\IP^1 \times \IP^1$, or $\IP^2$ blown up 
at $n$ points where $1\leq n \leq 8$.  
Our $\bX$
is a complex line bundle over $\calb$.   (In particular,
we take the canonical line bundle $\calo(K)$
over $\calb$.)

The D-branes 
in this geometry 
fill $\IR^{3,1}$ and wrap holomorphic cycles in $\bX$.
(We work with type IIB string theory.)
We assume there is a locus in the
Kaehler moduli space of $\bX$ where
$\calb$ shrinks to zero size.
At this point in the moduli space,
we expect  
an enhanced gauge symmetry for
D-branes wrapping cycles in $\calb$,
essentially because of all the strings
between the D-branes that become massless.
In the case of $\IP^2$, this locus of 
enhanced symmetry corresponds to the point
in moduli space where $\bX$ becomes
the orbifold $\IC^3/\IZ_3$ \cite{DFR}.
By an abuse of notation, we shall continue to call
this locus the ``orbifold point" for general
del Pezzos.  However, it is important to note that
in general the locus may contain more than one point and
$\bX$ may not be an orbifold on this locus.

%Instead of working with D-branes in $\bX$ directly, it is convenient
%to work with D-branes in the simpler del Pezzos.  We can then
%lift the D-branes back into $\bX$ with a minimal amount of work.

Kontsevich \cite{Kont} conjectured 
and later Douglas and collaborators \cite{DFR, dido, doug} 
gave further support to the idea that 
these holomorphically wrapped D-branes are objects in the
derived category of coherent sheaves $D^\flat(\bX)$.  
Bondal has shown that exceptional collections of coherent
sheaves on the del Pezzos generate $D^\flat(\calb)$
\cite{Bondal}.  In 
this esoteric mathematical context, it is not too 
surprising that exceptional collections should play
a vital role in understanding the low energy gauge theories
on these D-branes.  Indeed, Cachazo, Fiol, Intriligator, Katz, Vafa
\cite{CFIKV}, and later Wijnholt \cite{Wijnholt}
proposed a recipe for writing down a gauge theory
given an exceptional collection.  It is this recipe that
we analyze in detail using the lens of Seiberg duality.\footnote{
Exceptional collections have appeared elsewhere in the physics
literature in related contexts.  For example, they are
mirror duals of certain exceptional branes in Landau-Ginzburg
models \cite{zaslow, hiv}.}

%While the gauge theories dual to these del Pezzo geometries
%have several $SU(N)$ gauge groups and complicated
%matter content often described using quivers,
%Seiberg duality \cite{Seibergduality}
%was originally introduced as a way
%of understanding a far simpler gauge theory, 
%$\caln=1$ $SU(N)$ Yang-Mills with $N_f$ flavors.
%When $3N / 2 < N_f < 3N$, Seiberg noticed 
%there was a dual theory with $SU(N')$ gauge group,
%$N' = N_f - N$, and the same low
%energy degrees of freedom (see \cite{IntSei} for a review).
%Gauge group by gauge group, 
%it is straightforward to generalize this notion to the 
%more complicated quiver gauge theories of this paper.  

%Some of these questions were addressed by the author in \cite{Herzog}.
%We found a sufficient condition, christened well split, on the 
%gauge theory for Seiberg duality to correspond to a mutation.  Moreover,
 %for
%$\IP^1 \times \IP^1$ and $\IP^2$ blown up at a point,
%we proved that the well split property was preserved under Seiberg duality.
%sewellsplit, setheorem

In an earlier paper \cite{Herzog}, we found that Seiberg duality could
be defined in terms of mutations 
if the D-brane configuration (or
exceptional collection) was well split.  In order to have a consistent
definition, we conjectured in \cite{Herzog} that the Seiberg dual
of a well split quiver was well split.  

Taken together, the two principal results of this paper, Theorems \ref{sewellsplit}
and \ref{setheorem}, allow a consistent definition of Seiberg duality
in terms of mutations of strongly exceptional collections and prove
a weaker version of the well split conjecture.  
Theorem \ref{sewellsplit} identifies a subset
of well split exceptional collections, namely exceptional collections
which are not only strongly exceptional but which also generate something called a 
strong helix.  Theorem \ref{setheorem} states that the Seiberg dual of 
an exceptional collection which generates
a strong helix is another exceptional collection which also
generates a strong helix.  

These two results allow us consistently to ignore
the problematic partial Seiberg dualities mentioned above.
The types of gauge theories (not well split)
and exceptional collections (not strongly exceptional) that arise
from partial Seiberg dualities, never arise from performing
Seiberg duality.

Strong helices have appeared in the mathematics literature before.  Bondal \cite{Bondal}
calls them admissible helices. 
The condition that a D-brane configuration generates a 
strong helix is a statement about the kinds of open strings between
the D-branes (or equivalently $\Ext$ maps between
the sheaves in the exceptional collection).  

There is a separate more mathematical way of understanding our
results.
Bondal \cite{Bondal} defines admissible mutations to be those which
preserve the strongly exceptional property of the D-brane configuration.
Thus, we have demonstrated that Seiberg dualities can be understood
as admissible mutations.

To make the connection between well split and the strong
helix property, we introduce an intermediate notion, dubbed
$\Ext^{1,2}$ which, though technical, is a more obvious physical
constraint on the D-branes than strongly exceptional.
%The $\Ext^{1,2}$ condition is technical -- 
We will not give a precise definition of $\Ext^{1,2}$ until section 5, 
%The precise definition is 
%given above Lemma \ref{ext12} -- 
but we will be able partially to elucidate the physics
of it in section \ref{sec:bbrane}.
Two key auxiliary results in this paper are
1) Lemma  \ref{ext12} which states that
$\Ext^{1,2}$ implies the
collection is well split and 2) 
Propositions \ref{ext12impse} and \ref{seimpext12} which imply $\Ext^{1,2}$ is
equivalent to the strong helix property.  

\subsection{Reading Map}

This paper is meant to be read by both mathematicians and physicists.  As a result, there are
some sections which may be less interesting for one or the other half of the intended audience.
Section \ref{sec:bbrane} is intended primarily for a physics audience interested
in understanding why the derived category is useful for
studying gauge theory and is drawn largely from other sources
(see for example \cite{Aspinwall}).  Section \ref{sec:defs} sets up the exceptional
collection machinery with a level of rigor hopefully acceptable to
mathematicians and is also drawn from other sources (see for example
\cite{Rudakov}).  Section \ref{sec:gaugetheory}, which 
should be interesting to both physicists and mathematicians, describes how to convert
an exceptional collection on a del Pezzo into a quiver gauge theory, a recipe
worked out in \cite{CFIKV, Wijnholt, HW}.  Section \ref{sec:proof} contains the
proofs of the new results described in the introduction concerning the $\Ext^{1,2}$
and strongly exceptional
conditions.  The penultimate section contains a discussion of how our exceptional
collection definition of Seiberg duality matches the original gauge theory definition.

\section{Gauge Theories from B-Branes}
\label{sec:bbrane}

%The cartoon picture of how gauge theories arise from D-branes is very helpful, 
%but
%we will need to add more detail in order to understand Seiberg duality.
In this section, we make more precise the way in which gauge theories
arise from D-brane configurations.
The material below is drawn largely from other sources 
\cite{DFR, dido, doug, digo, mayr} (for a recent review, see for example 
\cite{Aspinwall}), but
we include it for coherence and completeness.

Our universe consists of D3-, D5-, and D7-branes 
in $\IR^{3,1} \times \bX$ where $\bX$ is a noncompact
Calabi-Yau three-fold.  We want to think of our gauge theory
as living in $\IR^{3,1}$, and so our D-branes span these
directions.  To preserve ${\mathcal N}=1$ supersymmetry,
we take the remaining transverse directions of the D-branes
to lie on holomorphic cycles in $\bX$.  If we think only in
terms of $\bX$, we have B-branes.  

The gauge theory is the low energy world-volume
description of the D-branes.  In particular, we only keep the massless
degrees of freedom, the degrees of freedom that morally at least correspond
to zero-length strings joining the D-branes.  But we need to be more precise.

First, we need a more precise definition of these D3-, D5-, and D7-branes.  Inside
$\bX$, they are submanifolds with some additional data.  
If we think of the D-brane intrinsically,
independent of its embedding, the additional data is a vector bundle.
However, once we put the D-brane back in its ambient space $\bX$,  we are
forced into using the language of coherent sheaves.

For example, consider the trivial line bundle ${\mathcal O}$.
Global sections of $\calo$ are holomorphic functions on $\bX$.  Consider also
${\mathcal O}(-D)$.  The global sections of $\calo(-D)$
have the analog of a single pole on the submanifold (or divisor) $D$.
There is clearly a map
\be
{\mathcal O(-D)} \rightarrow {\mathcal O}   \ ,
\label{cokernel}
\ee
which is given by multiplying sections of $\calo(-D)$ by
sections which vanish on $D$.
Morally, the cokernel of such a map, which we denote ${\mathcal O}_D$,
is a D-brane, i.e. the set of holomorphic functions on a submanifold (or
divisor) $D$.  However, such an object is not a vector bundle in $\bX$.  Mathematicians
would say that the kernel (or cokernel) 
of a map between vector bundles does not necessarily exist while the kernel of maps 
between coherent sheaves does.

As a first pass, then, D-branes correspond to coherent sheaves on $\bX$.
It turns out, however, that neither sheaves nor vector bundles are quite good enough.
Physically, we also need to be able to describe anti-branes.   
An anti-brane should correspond to an ``inverse'' sheaf with the opposite 
charges, but taking the inverse of a sheaf is difficult.
Instead, mathematicians introduce the notion of the derived category of coherent
sheaves, $D^\flat(\bX)$ where inverting is more natural.
In the derived category, sheaves get promoted to complexes that are zero everywhere
except at one position.  Let $\delta$ be the map that takes a sheaf $E$ into 
an object in $D^\flat(\bX)$:
\be
\delta E \equiv \cdots \to 0 \to E \to 0 \to \cdots \ .
\ee
Although the position of $E$ in a sequence has no meaning, once we introduce other
sheaves (or D-branes), we need to keep track of which elements in the sequence are nonzero.
Thus, $\delta E [n]$ is the complex with $E$ shifted $n$ places to the left.  In this
language, $\delta E$ shifted an odd number of places is an antibrane of $\delta E$.
To sum up, we have found that B-type D-branes are objects, i.e. complexes,
 in $D^\flat(\bX)$.

Next, we need a better definition of the massless degrees of freedom corresponding
to zero-length strings.  Massless degrees of freedom often come from zero modes of
a system, which are topological in nature, suggesting we look at the cohomology
of the D-branes.  Take two D7-branes corresponding to the line bundles
$\calo(E)$ and $\calo(F)$.  Someone truly inspired
might guess (correctly) that the massless degrees of freedom
should come from 
\be
H^{(0,k)}(\calo(F) \otimes \calo(E)^*)
\ee
for some $k$.  However, such a guess, while good for vector bundles, is insufficient for 
sheaves.  Luckily mathematicians have introduced a more general notion of cohomology
for sheaves called $\Ext$.  For vector bundles, we find
\be
H^{(0,k)}(\calo(F) \otimes \calo(E)^*) = \Ext^k(\calo(E), \calo(F)) \ .
\ee
On the three-fold $\bX$, our candidate massless degrees of freedom correspond to 
$\Ext^k$ where $k=0$, 1, 2, or 3.
There is a notion of Serre duality for $\Ext$ groups.  On 
$\bX$, Serre duality tells us that
\be
\Ext^k(E, F) = \Ext^{3-k}(F, E)^* \ .
\label{serreduality}
\ee

As a last step in identifying candidate massless degrees of freedom, we
need to lift $\Ext$ to $D^\flat(\bX)$, which fortunately 
has also been worked out by mathematicians.  One finds that
\be
\Ext^k(E,F) = \Hom^{k-p+q}_{D^\flat(\bX)} (\delta E [p], \delta F[q]) \ 
\ee
where no summation on $p$ and $q$ is implied.

Armed with a categorical description of D-branes and some candidates for massless
degrees of freedom, we need to check whether or not these degrees of freedom are
actually massless.  
The masses of the strings
depend on the Kaehler moduli space of $\bX$.  Strings that are massive at one
point in Kaehler moduli space may become massless or even tachyonic as we
move to another point!  In fact, more than the masses of the strings is at stake here.
A tachyonic string indicates the original pair of D-branes is unstable and will
condense to form a bound state.

The masses are most easily checked at the large volume
point in the Kaehler moduli space of $\bX$.  However, we will eventually need to 
compute the masses at the analog of the orbifold point.
Recall the case where $\bX$ is the complex line bundle $\calo(-3)$ over $\IP^2$.  The total
space is Calabi-Yau.  At the large volume point, the $\IP^2$ gets very large.
The orbifold point is $\IC^3 / \IZ_3$.  
For a general del Pezzo, 
at the analog of the orbifold point, $\calb$ should shrink to zero size
and we expect there to be extra massless open strings
which enhance the gauge symmetry of D-branes
wrapping cycles of $\calb$.

In general, we can define the D-brane charge of a sheaf $E$ 
\be
Z(E) = \int_\bX e^{-B-iJ} \ch(E) \sqrt{\Td(\bX)} + \ldots
\ee
where $\Td(\bX)$ is the Todd class of $\bX$, $B+iJ$ is
the complexified Kaehler form, and the $\ldots$
are instanton corrections which become more and more important
as we move away from large volume (large $J$). 
Fortunately, the Picard-Fuchs
equations and the mirror geometry usually allow one to calculate
$Z(E)$ numerically.  
We define the overall grading of the brane
\be
\xi(E) = \frac{1}{\pi} \arg Z(E) \mod 2 \ .
\ee
%At large volume, for a D$p$-brane, one finds that
%\be
%\xi(E) = -\frac{1}{2} \dim(p-3) \mod 2 \ .
%\ee

One can also define $Z$ and $\xi$ 
in the derived category.
At large volume, 
the categorical grading $n$ of $\delta E[n]$ multiplies
$Z(E)$ by $(-1)^n$:
\be
Z(\delta E[n]) = e^{i\pi n}   Z(E) \ .
\ee

The mass of the mode in $\Hom_{D^\flat(\bX)}^k(A,B)$ is given by
\be
M^2 = \frac{1}{2} ( \xi(B) - \xi(A) + k - 1) \ .
\label{mass}
\ee
The formula (\ref{mass}) is confusing in that $M^2$ changes as we move
away from the large volume point.  At large volume,
we might have a D7-brane anti-brane pair and a corresponding
tachyon with $M^2=-1/2$.  However, as we move to the ``orbifold point'',
instanton corrections can align the two $\xi$, leaving a massless field.
%This type of grading flow is critical for understanding the del Pezzo
%gauge theories we will encounter in later sections.

Now, we analyze the masses at the analog of the orbifold point.
At such a supersymmetric locus of points,
all the gradings
$\xi(A_i)$ of the branes should be aligned, i.e. $\xi(A_i) = \xi(A_j)$ $\forall i,j$
\cite{DFR, Aspinwall}.
We will assume in all that follows that such a supersymmetric locus of points
exists.  This assumption lets us ignore the Picard-Fuchs equations and
other subtleties of computing
$Z(E)$.

As a result of this assumption,
for $\Hom_{D^\flat}^k(A,B)$, 
$k=2$ or 3, the mass is positive and the corresponding degree of 
freedom should be absent in the low energy theory on the collection 
of D-branes at the ``orbifold point''.  
For $k=1$, we find a massless scalar mode.  For $k=0$, we find a
tachyon!  The ${\mathcal N}=1$ supersymmetry tells us that our fields should
fall into chiral or vector multiplets.

We recall a little elementary superstring theory and move 
beyond topological reasoning.  In quantizing
the string in flat space, we used the GSO projection to get rid of 
the tachyon.  In the present context, we can use the GSO projection to eliminate
the ground state string modes
corresponding to 
$\Hom_{D^\flat}^0(E,F)$ and $\Hom_{D^\flat}^2(E,F)$.  In this way, we eliminate all the tachyons.

Let's consider what kinds of fields are left.  Take $\Hom_{D^\flat}^0(E,E)$.
Although the ground state tachyon is projected out, just as in the more familiar context of 
quantizing superstrings in $\IR^{9,1}$, there is an excited massless string
state corresponding to a gauge boson (and, filling out the $\caln =1$ multiplet,
a gaugino).  
For $\Hom_{D^\flat}^1(E,E)$, we get adjoint
matter fields.  For $\Hom_{D^\flat}^1(E,F)$, we get matter fields transforming in the 
fundamental of one gauge group and the antifundamental of another.

On the other hand, $\Hom_{D^\flat}^0(E,F)$, $E \neq F$, is problematic.  
Although the ground state is
projected out, the first excited state is a gauge field transforming in the 
bifundamental 
representation of the gauge groups associated with $E$ and $F$.
Gauge fields by definition should transform in the adjoint representation of 
a Lie group.  Moreover, this Lie group should be a direct
product of simple Lie groups and $U(1)$ factors.
The gauge groups associated with $E$ and $F$
will be of type $SU(N)$.  The bifundamental representation of $SU(N)\times
SU(M)$ is the adjoint representation of some Lie group, but that Lie group does
not have this direct product structure.\footnote{I would like to thank 
D. Berenstein
for this argument.}
(See however \cite{VafaMN} for a different perspective on these fields.)  
%The reader is invited to try to write down
%a consistent kinetic term in the Lagrangian for such a field.
It is precisely these types of fields that can arise under the partial Seiberg
dualities discussed in the introduction.

We are now in a position to elucidate in part the $\Ext^{1,2}$ condition.  
One consequence of the $\Ext^{1,2}$ condition
is that for $E \neq F$, $k$ in $\Hom^k_{D^\flat}(E,F)$ 
is either one or two.  Thus, these weird bifundamental gauge fields will
never appear and will never be produced by Seiberg duality.
In other words, we can consistently ignore all the D-brane 
configurations which give rise to bifundamental gauge fields.

\section{Exceptional Sheaves on del Pezzos}
\label{sec:defs}

We are interested in exceptional collections of
coherent sheaves $E$ supported on a del Pezzo 
surface $\calb$.  The standard mathematical reference, from
which we draw much of this section, is \cite{Rudakov}.

\begin{definition}
A sheaf $E$ is called {\it exceptional} if $\Hom(E,E) \cong \IC$ and 
$\Ext^i(E,E) = 0$ for $i>0$.
\end{definition}

Recall that $\Hom \equiv \Ext^0$.
In gauge theory language, we would associate a gauge group to each sheaf.
Exceptional sheaves have no adjoint matter.
In \cite{KO} it is proven that an exceptional sheaf $E$ 
on a del Pezzo is either a vector bundle or a torsion sheaf of the form
$\calo_\ell(m)$ where $\ell$ is an exceptional curve.

\begin{definition}
An {\it exceptional collection} is an ordered set of exceptional
sheaves $(E_1, E_2, \ldots, E_n)$ with the following additional properties:
\begin{enumerate}
\item{$\Ext^k(E_i, E_j) = 0$  $\forall k$ if $i>j$;}
\item{$\Ext^k(E_i, E_j) = 0$ for all but at most one $k$ if $i<j$.}
\end{enumerate}
\end{definition}

For gauge theories, the $\Ext$ maps between sheaves correspond to
bifundamental matter fields.  The exceptional collection condition implies
that the matter associated with any particular sheaf/gauge group is chiral.

\begin{definition}
A {\it strongly exceptional collection} is an exceptional collection
such that only $\Hom(E_i,E_j)$ is nonzero. 
\end{definition}

Kuleshov and Orlov (see section 2.11 of \cite{KO}) 
demonstrated that for del Pezzo surfaces, for an
exceptional pair $(E,F)$, the maps
$\Ext^i(E,F)$ can be non-zero for either only $\Hom$ or
only $\Ext^1$.

\subsection{Euler Character}

We will gradually see the significance of strong exceptionality,
but we remark in the meantime that the Euler character 
$\chi(E,F)$ makes
exceptional collections relatively easy to work with:
\be
\chi(E,F) \equiv \sum_k (-1)^k \dim \Ext^k (E,F) \ . 
\ee
In particular, the Euler character
is an efficient tool for calculating $\dim \Ext^i (E,F)$ for an exceptional
pair $(E,F)$ (and hence the number of bifundamentals in the gauge theory).
Riemann-Roch implies that
\begin{eqnarray*}
\chi(E,F) &=& \int_{\calb} \ch(E^*) \ch(F) \Td(\calb) \\
&=& r(E) r(F) + \frac{1}{2}( r(E) \deg(F) - r(F) \deg(E) )\\
&& +r(E) \ch_2(F) + r(F) \ch_2(E) - c_1(E)\cdot c_1(F) \ 
\end{eqnarray*}
where we have used the Chern character of the sheaf 
\be
\ch(E) = (r(E), c_1(E), \ch_2(E)) \ .
\ee
Also, $\Td(\calb) = 1 - \frac{K}{2} + H^2$,
where $K$ is the canonical class and $H$ is the hyperplane, with
$\int_\calb H^2 = 1$.  Finally, the degree $\deg(E) = (-K) \cdot c_1(E)$.
\begin{definition}
We define the slope of a torsion free sheaf $E$ to be
$\mu(E) = \deg(E)/ r(E)$.
\end{definition}
The slope of a torsion sheaf $\mu(\calo_\ell(m))$ is taken to be
infinite.

We denote $\chi_-(E,F)$ to be the antisymmetric part of $\chi$
\be
\chi_-(E,F) = r(E) \deg(F) - r(F) \deg(E) \ .
\ee
Note that for an exceptional pair $(E,F)$, $\chi_-(E,F) = \chi(E,F)$.
For torsion free sheaves, we may write
\be
\chi_-(E,F) = r(E) r(F) (\mu(F) - \mu(E)) \ .
\ee

\subsection{Derived Category}

We denote by $D^\flat(\calb)$ the bounded derived category of coherent
sheaves on $\calb$.  
The main reason for moving to the derived category is that we will
need a notion of an inverse exceptional collection, and while sheaves
do not necessarily have well defined inverses, their images in the derived category do.

We can map coherent sheaves $E$ to objects in 
the derived category in a simple way.  In particular, we denote by
$\delta E$ the complex in $D^\flat(\calb)$ which has only one nonzero entry,
$E$.  We are free to shift the location of $E$ in the complex by $k$ units to the left,
which we denote $\delta E[k]$.  Note that
\be
\Hom_{D^\flat}^k(\delta E[i], \delta F[j]) = \Hom^{k-i+j}_{D^\flat}(\delta E, \delta F) =
\Ext^{k-i+j}(E,F) \ .
\ee

The notion of exceptionality extends naturally to $D^\flat(\calb)$.  
An object $A \in D^\flat(\calb)$ is called {\it exceptional} if
$\Hom^0_{D^\flat(\calb)}(A,A) \cong \IC$ and
$\Hom^i_{D^\flat(\calb)}(A,A) = 0$ for $i\neq 0$.  The notion of exceptional
collection generalizes as well. For a set of objects $(A_1, A_2, \ldots A_n)$,
$\Hom^k_{D^\flat(\calb)}(A_i, A_j) = 0$ for $i>j$ and
$\Hom^k_{D^\flat(\calb)}(A_j, A_i) = 0$ for all but at most one $k$.

We define the Euler character for complexes with one nonzero element in the following
way:
\be
\chi(\delta E[j], \delta F[k]) = (-1)^{k-j} \chi(E,F) \ .
\ee

\subsection{Mutations}

A useful way of generating new exceptional collections from old ones is
mutation.  
We will eventually see that certain sequences of mutations are
Seiberg dualities.
Let $(A,B)$ be an exceptional pair of objects in $D^\flat(\calb)$.
There exists a canonical morphism
$\RHom(A,B) \otimes A \stackrel{can}{\to} B$.  Let 
the {\it left mutation} of $B$ over $A$, denoted $L_A^D B$, be the object
which completes this morphism to a distinguished triangle
\be
L_A^D B[-1] \to \RHom(A,B) \otimes A \stackrel{can}{\to} B \to L_A^D B \ .
\label{dtril}
\ee
Equivalently, one may define {\it right mutation} as the object which
completes the canonical morphism $A \to \RHom^*(A,B) \otimes B$ to
the distinguished triangle
\be
R_B^D A \to A \to \RHom^*(A,B) \otimes B \to R_B^D A[1] \ .
\ee
\begin{theorem}[see for example Section 2.3 of \cite{Gorodentsev}]
If $(A,B)$ is an exceptional pair in $D^\flat(\calb)$, then 
$(L_A^D B, A)$ 
and $(B, R_B^D A)$ are exceptional pairs in $D^\flat(\calb)$.
\end{theorem}

For a pair $(A,B) = (\delta E, \delta F)$, 
the distinguished triangle (\ref{dtril}) will reduce to a short exact
sequence of sheaves.  If $\Ext^1(E,F) \neq 0$, the mutation is called
an extension and takes the form
\be
0 \to F \to L_E F \to \Ext^1(E,F) \otimes E \to 0 \ .
\ee  
If $\Hom(E,F) \neq 0$, we need to make sure that $L_E F$ is
defined in such a way that 
$r(L_E F) = \pm ( r(F) - \chi(E,F) r(E)) \geq 0$.  There are two
possibilities: 
\begin{eqnarray*}
0 \to L_E F \to \Hom(E,F) \otimes E \to F \to 0 && \mbox{(division)} \ , \\
0 \to \Hom(E,F) \otimes E \to F \to L_E F \to 0 && \mbox{(recoil)} \ .
\end{eqnarray*}

Similarly for right mutation, we find the three short exact sequences
\begin{eqnarray*}
0 \to E \to \Hom(E,F)^* \otimes F \to R_F E \to 0 && \mbox{(division)} \ , \\
0 \to R_F E \to E \to \Hom(E,F)^* \otimes F \to 0 && \mbox{(recoil)}  \ ,\\
0 \to \Ext^1(E,F)^* \otimes F \to R_F E \to E \to 0 && \mbox{(extension)} \ .
\end{eqnarray*}

In the case where $L_F E$ or $R_F E$ is torsion, there is an ambiguity in
the positive rank prescription which we now eliminate.  Note from our elementary
discussion of torsion sheaves around (\ref{cokernel}) that we expect these torsion 
sheaves
to show up as cokernels in short exact sequences.  Thus, if $L_F E$ is torsion,
we conclude the sequence must be a recoil while if $R_F E$ is torsion, then
the sequence is a division.

Now we come to a crucial observation about the relation between mutation
in $D^\flat(\calb)$ and mutation of coherent sheaves.

\begin{proposition}[Proposition 1.8 of \cite{NK}]
If a sheaf mutation of a pair $(E,F)$ is either a recoil or an extension, then
$L_{\delta E}^D \delta F = \delta L_E F$ and $R_{\delta F}^D \delta E =\delta R_F E$.
For the case of division, $L_{\delta E}^D \delta F = \delta L_E F[1]$ and
$R_{\delta F}^D \delta E = \delta R_F E [-1]$.
\end{proposition}

\subsection{Braid Group}

One can think of mutations as an action of the braid group on an
exceptional collection.

We define the following maps $L_i$ and $R_i$ on an exceptional 
collection $\cale = (E_1, \ldots, E_n)$:
\begin{eqnarray*}
L_i : (\ldots, E_{i-1}, E_i, E_{i+1}, \ldots) &\to &
(\ldots, E_{i-1}, L_{E_i} E_{i+1}, E_i, \ldots) \ , \\
R_i: (\ldots, E_{i-1}, E_i, E_{i+1}, \ldots) &\to &
(\ldots, E_{i-1}, E_{i+1}, R_{E_{i+1}} E_i, \ldots ) \ .
\end{eqnarray*}

\begin{proposition}[Section 2.3 of \cite{Gorodentsev}, Assertion 2.3 of
\cite{Bondal}]
The mutations $R_i$ and $L_i$ are inverse, $R_i \circ L_i = \id$.
Moreover, right and left mutations define an action of the $n$-string
braid group on $\cale$:
\be
R_i \circ R_{i+1} \circ R_i = R_{i+1} \circ R_i \circ R_{i+1} \; , \; \; \; 
L_i \circ L_{i+1}  \circ L_i = L_{i+1} \circ L_i \circ L_{i+1} \ .
\label{braidrels}
\ee
\end{proposition}

\subsection{Helices}

A {\it helix} $\calh = (E_i)_{i\in \IZ}$ is a bi-infinite extension of an 
exceptional collection $\cale$ defined recursively by
\begin{eqnarray*}
E_{i+n} &=& R_{E_{i+n-1}} \cdots R_{E_{i+1}} E_i \ , \\
E_{-i} &=& L_{E_{-i+1}} \cdots L_{E_{n-1-i}} E_{n-i} \; \; \; i \geq 0 \ .
\end{eqnarray*}
such that the helix has period $n$, by which we mean
\be
E_i = E_{n+i} \otimes K \; \; \; \; \forall i \in \IZ \ .
\label{helixperiod}
\ee
A {\it foundation of a helix} is any subcollection of the form
$(E_{m+1}, \ldots, E_{m+n})$, where $m \in \IZ$.
An exceptional collection is called {\it complete} if it generates
$D^\flat(\calb)$.
\begin{theorem}[Theorem 4.1 of \cite{Bondal}]
An exceptional
collection is complete if and only if it is the foundation of a helix. 
\end{theorem}

A corollary that can be extracted from Bondal's proof of the preceding
theorem (and Serre duality) is that
\be
L^D_{\delta E_{1}} L^D_{\delta E_{2}} \cdots L^D_{\delta E_{n-1}} \delta E_n = 
\delta (E_n \otimes K) [2] \ .
\label{leftshift2}
\ee

Helices provide us with a stronger notion of strongly exceptional which
will be important for our proofs.
\begin{definition}
A {\it strong helix} is a helix $\calh$ such that every foundation $\cale$ of $\calh$
is strongly exceptional.
\end{definition}
\begin{remark}\label{strslopes}
Given an exceptional collection  
$\cale = (E_1, \ldots, E_n)$, $\cale$ generates a strong helix
if and only if the slopes satisfy
\be
\mu(E_1) \leq \mu(E_2) \leq \cdots \leq \mu(E_n) \leq \mu(E_1) + K^2 \ .
\ee
\end{remark}  
The remark follows from (\ref{helixperiod}).  
In the forward direction, we find that $\cale$ must be strongly exceptional.
Therefore, the first $n-1$ inequalities follow.  We could also choose
a different foundation which included $E_n \otimes K$ and $E_1$ as neighbors
in the collection.  Then the last inequality follows because tensoring
with $K$ subtracts $K^2$ from the slope of the sheaf.
In the reverse direction, we know that all foundations of the helix
can be generated by tensoring the sheaves in $\cale$ with
$K$ the appropriate number of times.  
Given any foundation, $n-1$ of the $n$
inequalities above are sufficient to guarantee that the
foundation is strongly exceptional.
(Note we have departed from the conventions of Bondal \cite{Bondal} here,
who calls such helices admissible.)

\section{Quiver Gauge Theories}
\label{sec:gaugetheory}

\begin{definition}
A {\it quiver} is a collection of nodes (or vector spaces) and arrows (or maps between
the vector spaces).
\end{definition}

For nodes $i$ and $j$, we denote a map from vector space 
$V_i$ to $V_j$ as $X_{ij}$.
A {\it quiver with relations} is a quiver where certain compositions
of maps are identified.

\begin{definition}
A ${\mathcal N}=1$ {\it supersymmetric quiver}  is a quiver with
relations such that all the relations are generated by a 
superpotential $W$
and are of the form
\be
\frac{\partial W}{\partial X_{ij}} = 0 \ .
\ee
Moreover, the superpotential must be a polynomial in the $X_{ij}$
and correspond to a sum over loops in the quiver.
\end{definition}

To each exceptional collection of sheaves $\cale = (E_1, \ldots, E_n)$, 
we can associate a supersymmetric quiver in the following way.
To calculate the number of
bifundamentals or maps between the nodes, we first produce the left-dual collection
or gauge theory collection
\be
\cale^Q = (E_n^\vee, \ldots, E_1^\vee) = 
(L^D_{\delta E_1}\cdots L^D_{ \delta E_{n-1}} \delta E_n, \ldots, L^D_{\delta E_1} \delta E_2, 
\delta E_1) \ .
\label{dualdef}
\ee
Note that this collection is dual in the sense of the Euler character:
\be
\chi(E_i, E_j^\vee) = \delta_{ij} \ .
\label{eulerdual}
\ee
We define the upper triangular matrix 
\be
S_{ij} = \chi(E^\vee_j, E^\vee_i) \ .
\ee
Note that $\cale^Q$ lives in $D^\flat(\calb)$ while $\cale$ is a collection
of sheaves.
The number of arrows from node $i$ to node $j$ is defined to 
be $S_{ij}$ for $i \neq j$, where $S_{ij} < 0$ implies we reverse
the direction of the arrows.
We will call a quiver produced in this way an {\it exceptional quiver}.
At the moment, it is only known how to produce cubic superpotential
terms from $\cale$ \cite{Wijnholt}.
%\footnote{
%Hanany paper on exceptional groups?}

Note that the original or geometric collection $\cale$ generates a helix
$\calh$ in the usual way.  One might worry that different foundations of
the helix produce different quivers.

\begin{theorem}[see section 3 of \cite{Herzog}]
The quiver produced from a helix $\calh$ is independent of
the particular foundation $\cale$ chosen.
\end{theorem}

This construction can be partially motivated for strongly exceptional
collections in the following way.  We begin by considering the
algebra of homomorphisms.
\be
{\mathcal A} = \bigoplus_{i,j=1}^n \Hom(E_i, E_j)
\ee
on the del Pezzo.
Note that for ordinary exceptional collections, $\Ext$ maps would
appear which would make the interpretation of this algebra
more painful.

To construct the supersymmetric quiver from $\cala$, we make an
intermediate step of constructing a quiver whose path algebra is
the same as $\cala$, the so called Beilinson quiver.  
For the Beilinson quiver, $\dim \Hom (E_i, E_j)$ has an interpretation
as the total number of paths between nodes $i$ and $j$ up to relations.
In constructing the quiver, we cannot put $\dim \Hom(E_i, E_j)$ arrows
between nodes $i$ and $j$ unless $j = i+1$ because then we would
have far too many paths.  We can only write down the generators of the
algebra.

Roughly,
the negative entries of $S_{ij}$ correspond to generators of the 
algebra ${\mathcal A}$ while the positive entries correspond
to relations.  
$S$ is an upper triangular matrix with ones along the diagonal.
Thus $S^{-1}$ where 
$S_{ij}^{-1}  = 
\dim \Hom(E_i, E_j)$ is also upper triangular with ones along the diagonal.
The claim follows from writing $\dim \Hom(E_i, E_j)$ in terms of the individual elements of
$S$.
However, we say roughly because in general there can be higher relations: relations between
relations and so forth.  These higher order relations do not occur once we restrict
to strong helices in section 5.

To get the supersymmetric quiver from the Beilinson quiver, we promote relations to generators.
By assumption, the relations in the supersymmetric quiver come from a superpotential $W$.
When a relation becomes a generator $X_{ij}$ in moving from the Beilinson quiver to the supersymmetric 
quiver, the old relation can be represented as $\partial W / \partial X_{ij} = 0$.

We leave a physics motivation to the end of subsection \ref{ssec:quivreps}

\subsection{Representations of the Quiver}
\label{ssec:quivreps}

We get representations of the quiver by choosing ranks $d^i$ for
the vector spaces $V_i$.   To each representation, we associate 
an object $A$ in the derived category (although here we really 
only need the chern characters):
\be
\ch(A)  = \sum_{i=1}^n d^i \ch(E^\vee_i) \ .
\ee
Certain special representations are singled out, the representations which
satisfy chiral anomaly cancellation
\be
(S - S^T) \cdot d = 0 \ .
\ee
The relation is satisfied for objects $A$ 
such that $\chi(E_i^\vee, A) = \chi(A, E_i^\vee)$ $\forall i$.
Such objects have $r(A)=0$ and $\deg(A) = 
(-K)\cdot c_1(A)=0$.
One such object corresponds to the skyscraper sheaf at a point,
$\delta {\mathcal O}_{pt}$.
For each exceptional divisor $\ell_I$ in the del Pezzo, we
construct an additional object satisfying chiral anomaly cancellation.
For a del Pezzo with three or more exceptional divisors,
one could choose, for example, objects with chern characters
$\ch(A_I) = (0,\ell_1 - \ell_I, 0)$ for $I>1$ and $\ch(A_1) = 
(0, H- \ell_1 - \ell_2 - \ell_3, 0)$.

If $d$ satisfies anomaly
cancellation, then it must be of the form
\be
d^i = r^i N + s^i_I M^I
\ee
where $r^i = r(E_i)$, $s^i_I = c_1(A_I) \cdot c_1(E_i)$, and $N$ and the $M^I$
are arbitrary integers.  
%For each node, we get a gauge group $SU(d^i)$
%in the gauge theory. 
In D-brane language, ${\mathcal O}_{pt}$ corresponds to a D3-brane
while the remaining $A_i$ correspond to D5-branes wrapped on the
exceptional divisors $\ell_I$.

\begin{example}\label{Ptwo}
Let us consider the collection $\cale = ({\mathcal O}, {\mathcal O}(1), {\mathcal O}(2))$
on $\IP^2$.  First we construct 
$\cale^Q = (\delta {\mathcal O}(-1)[2], \delta T^*(1)[1], \delta {\mathcal O})$
where $T^*(1)$ is the twisted cotangent bundle defined by the short exact sequence
(division)
\be
0 \to T^*(1) \to \IC^3 \otimes {\mathcal O} \to {\mathcal O}(1) \to 0 \ .
\ee
Note that 
\be
S = 
\left(
\begin{array}{rrr}
1 & -3 & 3 \\
0 & 1 & -3 \\
0 & 0 & 1 
\end{array}
\right)
\ee
and the kernel of $S-S^T$ is one dimensional; there are no $s^i_I$ in this 
example.  The vector $d$ is a multiple of $(1,1,1)$.
\end{example}

A precise definition of a supersymmetric field theory is
difficult to present and also not so important for the
following.  However, this work was inspired by the 
desire to use exceptional collection techniques
to understand field theories.  Thus, 
to each supersymmetric quiver, we may associate
an ${\mathcal N}=1$ {\it supersymmetric quiver gauge theory}.  
In particular,
for every node, we have a $SU(d^i)$ gauge group.  For every
map $X_{ij}$, we have a bifundamental chiral superfield
transforming in the fundamental representation of $SU(d^i)$ and the
antifundamental of $SU(d^j)$.  Quiver invariants such as
$\Tr X_{ij} X_{jk} X_{ki}$, where the trace is over the $SU(d^i)$
indices, become gauge invariant operators in the field theory
and are important to physicists.

We come now to a physical motivation for
the construction of a supersymmetric quiver, 
at least for the $\Ext^{1,2}$ configurations mentioned in the introduction.

The gauge theory arises from a D-brane configuration
in the total space $\bX$ of the 
complex line bundle over the 
del Pezzo.  Consequently, we must lift our sheaves up into $\bX$.  
We will take the simplest approach and ``extend by zero".  In other words,
our sheaves only have support on the
submanifold $\calb \in \bX$.
Away from $\calb$, the sections of our sheaves vanish.  If $E$ is a sheaf
on $\calb$, let its extension by zero be denoted $\tilde E$.

By a result of Seidel and Thomas \cite{SeiTho} and also of \cite{CFIKV},
\be
\Ext^i(\tilde E, \tilde F) = \Ext^i(E,F) \oplus \Ext^{3-i} (F, E)^* \ .
\ee 
Essentially, one finds that the two $\Ext$'s are the same up to 
the modification needed for consistency with Serre duality
(\ref{serreduality}).  This result can then be adjusted to account for the gradings of the
derived category:
\be
\Hom_{D^\flat(\bX)}^i(\tilde A, \tilde B) = 
\Hom_{D^\flat(\calb)}^i(A,B) \oplus \Hom_{D^\flat(\calb)}^{3-i} (B, A)^* \ .
\ee
where we can assume if necessary that $A$ and $B$ are complexes with only one
non-zero entry.

In section \ref{sec:bbrane}, the massless bifundamental matter fields corresponded to nonzero
\be
\Hom_{D^\flat(\bX)}^k(A,B)
\ee
 with $k=1$.  
A nonzero $\Hom_{D^\flat(\calb)}^k$ with $k=1$ or $k=2$
will lift to a nonzero $\Hom_{D^\flat(\bX)}^k$ with $k=1$ and $k=2$.
The GSO projection eliminates the $k=2$ state in $\bX$.
(Also, the $k=2$ state is massive and absent at low energies.)
A $k=2$ state in $\calb$ will become a $k=1$ state in $\bX$ with
the oppositive orientation, i.e. we would draw the arrow in the opposite direction
in the quiver.

Thus in the case where $\cale^Q$ contains nonzero $\Hom_{D^\flat(\calb)}^k(E_i^\vee, E_j^\vee)$,
$i \neq j$,
only for $k=1$ and $k=2$, the prescription given above for writing down a quiver
based on the sign of the $S_{ij}$ matches the discussion in section \ref{sec:bbrane}
concerning bifundamental chiral fields.

To review, we start with $\IR^{3,1} \times \bX$ and some D-branes wrapped
holomorphically on $\calb \in \bX$.  We assume that there
is an ``orbifold point'' (or perhaps locus) in the Kaehler moduli space of
$\bX$ where the gauge symmetry on the D-branes is enhanced.  At this orbifold
point, we assume that an $\Ext^{1,2}$ exceptional collection on $\calb$
provides a complete collection of mutually supersymmetric fractional 
branes.  The collection is complete in the sense that the K-theory charges
(or chern character) of any D-brane can be represented as a sum over the
K-theory charges of the fractional branes.  
Because of anomaly cancellation considerations,
we only consider D-brane configurations with the same K-theory
charges (or chern characters) as a collection of D3-branes and D5-branes
wrapped on the exceptional divisors of $\calb$.

\section{Seiberg Duality}
\label{sec:proof}

We would like to define Seiberg duality as an action on
$\cale$.  As originally defined, Seiberg duality is a
transformation on an ${\mathcal N}=1$ gauge theory.
We will see how our definition matches the usual
definition in section \ref{sec:usualdef}.
We do not know how to define 
Seiberg duality on a generic $\cale$.  We must first
introduce the notion of {\it well split}.

\begin{definition}
A node $i$ of an exceptional quiver is {\it well split} if we can choose a foundation
of $\calh$ such that for all $j<i$, the arrows are outgoing from $i$ while for
all $j>i$, the arrows are ingoing into $i$. 
\end{definition}

\begin{figure}
\includegraphics[width=3in]{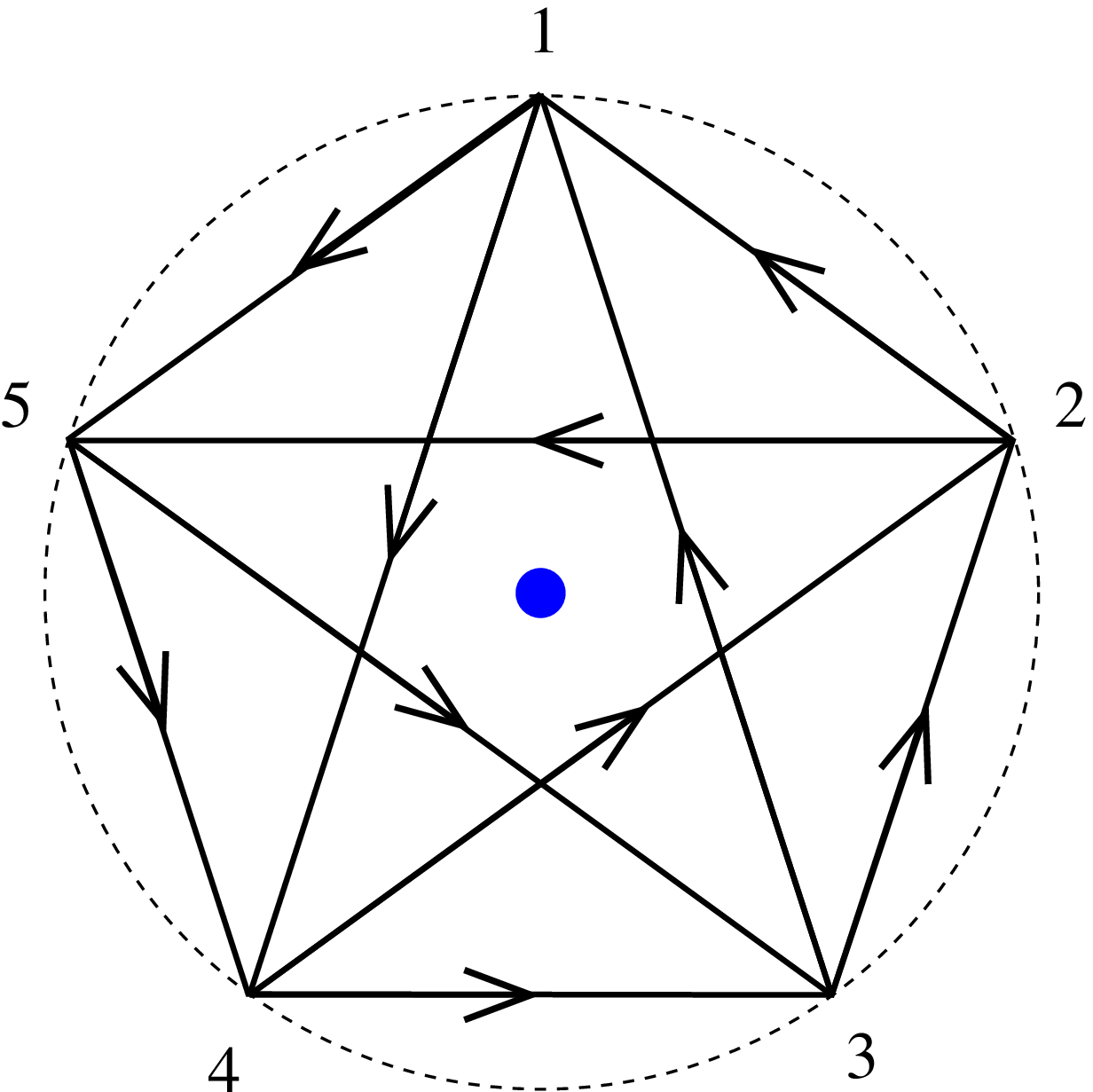}
\hfil \hfil
\includegraphics[width=3in]{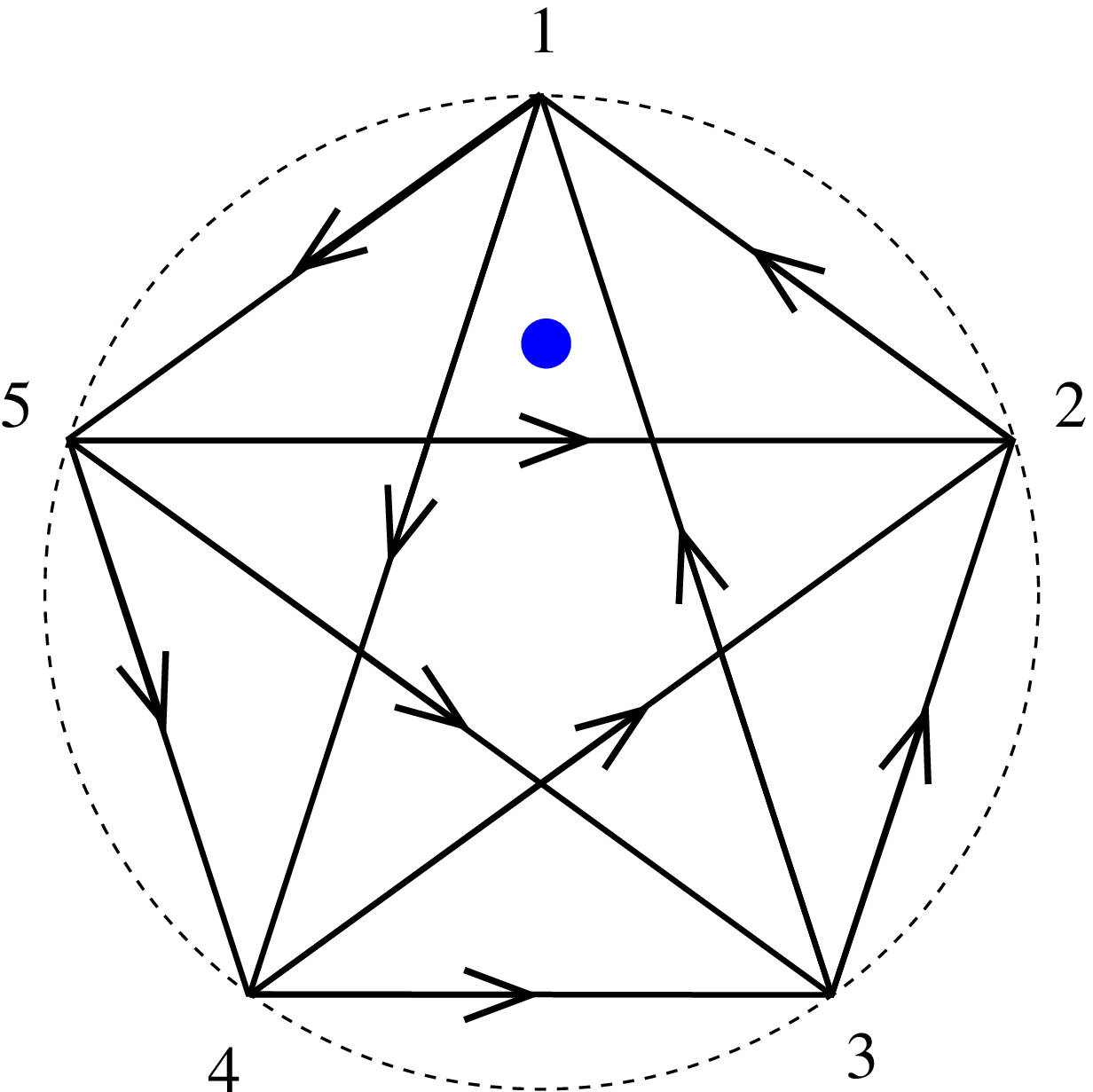}
\vfil
\vfil
\begin{center}
\includegraphics[width=3in]{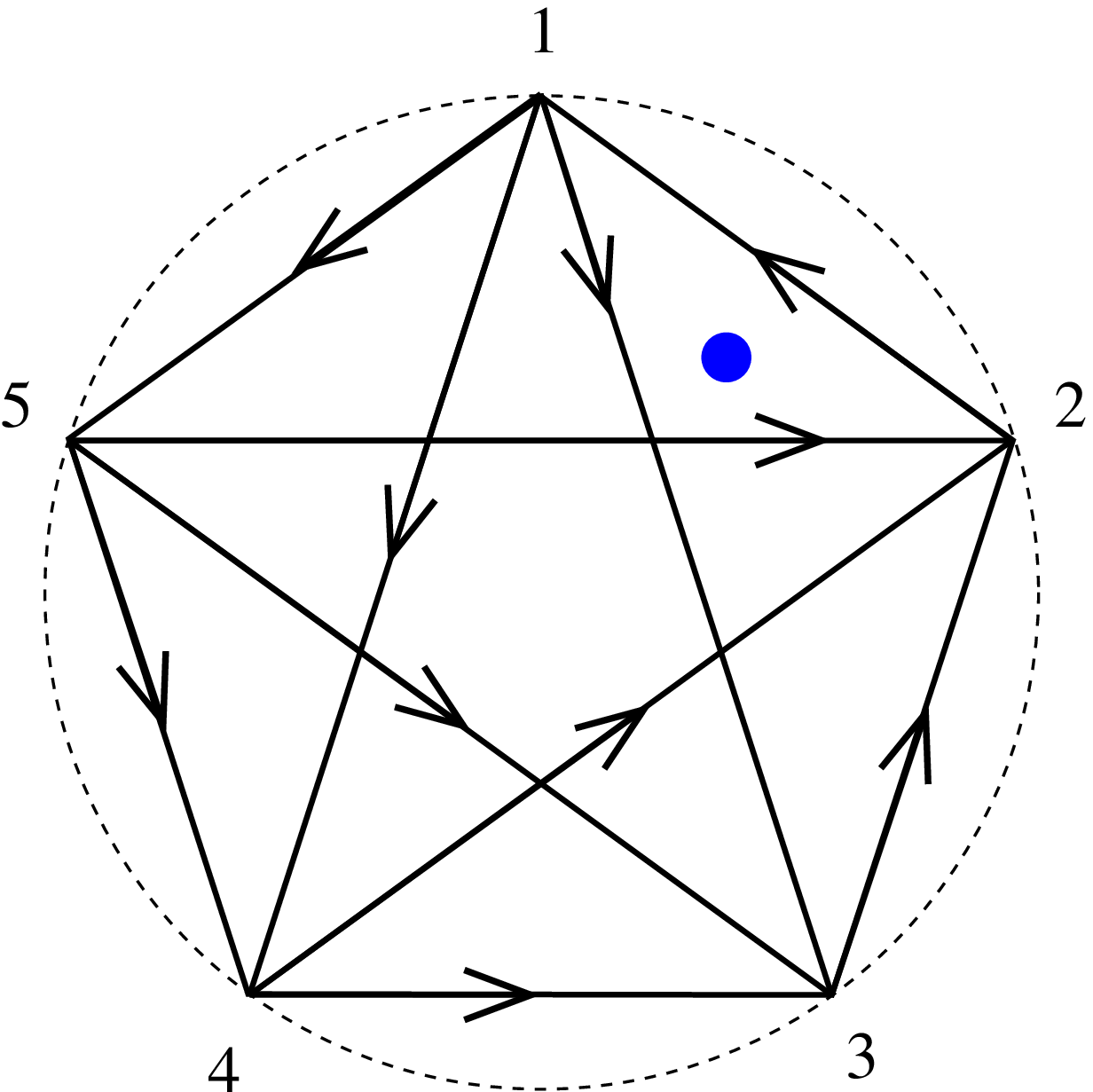}
\end{center}
\caption{Well split, five node quivers.  The blue dot identifies a special
polygon around which all arrows travel counterclockwise.}
\label{pentfig}
\end{figure}

Well split is perhaps more easily thought of pictorially 
in terms of the quiver.  We order the nodes of the quiver
on a circle as they appear in the foundation of the helix.  We then
construct the quiver as described in section 4.  If a node is well split,
we can draw a line through the node that divides the quiver into
two pieces such that the arrows incident on the node satisfy a special
property.  In particular, in one piece, all these arrows will be
incoming to the well split node.  In the other piece, the arrows will
all be outgoing from the well split node.  In Figure \ref{pentfig}, 
the extra line that divides the quiver into two pieces joins the node and the
blue dot.

We can now define the action of Seiberg duality, $SD$, on a well split node.
The idea is to left mutate (or right mutate), the well split node $i$ past all
the outgoing (respectively ingoing) nodes. 
\be
SD : (E_1, \ldots, E_i, \ldots E_n) \to (L_{E_1} \cdots L_{E_{i-1}} E_i, 
E_1, \ldots,  E_{i-1}, E_{i+1}, \ldots, E_n) \ .
\ee
From the helix property, we
know the result is independent of whether we perform right or left mutation.

Clearly, this definition of Seiberg duality would be more useful if it could be performed
at every node of a quiver.  We are naturally led to consider {\it well split quivers}.
\begin{definition}
A quiver is well split if every node is well split.
\end{definition}

An interesting subset of well split quivers satisfies a different pictorial constraint.
Instead of checking each node individually that all the incoming arrows come from the
right in the collection and all the outgoing arrows go to the left, 
one searches the quiver for a polygon such that around the polygon, all
the arrows travel counterclockwise.\footnote{I would like to thank
Paul Aspinwall for this suggestion.}  Note that
the quiver may have to be deformed to get the polygon to appear.

Figure \ref{pentfig} shows the three types of well split five node quivers that typically
appear.  (Note there are also degenerate cases where some arrows are missing.)
The three types correspond to the three types of polygons in this geometric 
construction.

\begin{figure}
\begin{center}
\includegraphics[width=5in]{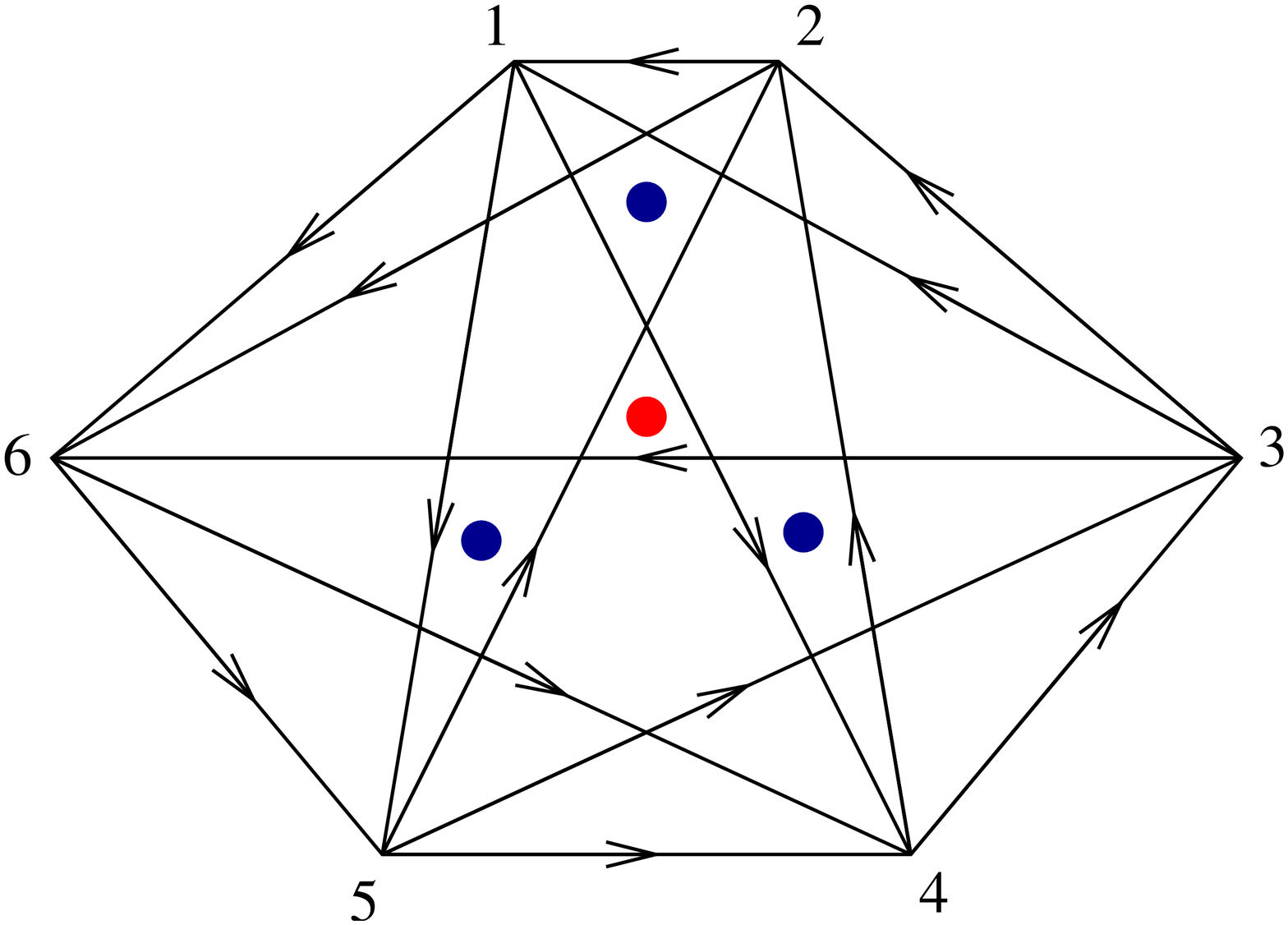}
\end{center}
\caption{A well split six node quiver.  The blue dots label polygons around which
only nearby arrows travel counterclockwise.  The red dot labels a triangle which
can disappear if the nodes are moved.  Indeed, if we move nodes 1 and 2
farther apart and 4 and 5 closer together, the triangle with a red dot will 
disappear and a new triangle will take its place around which {\it all} arrows
travel counterclockwise.}
\label{sixfig}
\end{figure}

The existence of such a polygon clearly implies well split.  Put a point inside the
polygon.  Pick any node.  Join the point and the node with a line.  
All the arrows incident on this node 
to one side of this line will be ingoing while all the arrows on the other side will be outgoing.
That well split implies the existence of such a polygon is not trivial, and we leave it as a conjecture. Figure \ref{sixfig} provides an apparent counterexample.  However,
we can deform the quiver such that the triangle with a red dot disappears and a new
triangle appears with the required counterclockwise property.

Exploring the value of our sheaf theoretic
definition of Seiberg duality, 
we are also led to wonder if we can take arbitrary Seiberg duals of Seiberg duals.
For that we would need the following.
\begin{conjecture}[Section 5 of \cite{Herzog}]
The Seiberg dual of a well split quiver is well split.
\end{conjecture}
The conjecture is trivial for three node exceptional 
quivers and was proved
for the four node exceptional quivers in \cite{Herzog}.
In the following, we will prove what superficially appears to be a weaker
version of this conjecture.  However, we are uncertain that a stronger version
exists.

Our two principal results are
\begin{theorem} 
\label{sewellsplit}
A strong helix $\calh$ generates a well split
quiver.
\end{theorem}
and
\begin{theorem}
\label{setheorem}
Let $\calh$ be a strong helix.
Let $\cale$ be a foundation of $\calh$.  
The Seiberg dual $\cale'$ of $\cale$ 
generates a strong helix
$\calh'$.
\end{theorem}

\subsection{Proof of Theorem \ref{sewellsplit}}

To prove Theorem \ref{sewellsplit}, we introduce an intermediate notion,
$\Ext^{1,2}$.  We will prove that $\Ext^{1,2}$ implies
well split and then finish by demonstrating
that $\Ext^{1,2}$ is equivalent
to the strong helix property.

In this paper, we are concerned with $\cale$ constructed
from del Pezzo surfaces.  We see from (\ref{leftshift2})
that the object $E_n^\vee \in \cale^Q$ 
considered as a sheaf will have a shift of grading
equal to two.  Meanwhile, the object 
$E_1^\vee = \delta E_1$
is not shifted at all.  Because the shifts of grading can 
only increase as we left mutate $E_i$, the non-zero sheaf
in the complex $E_i^\vee$
can be shifted only zero, one, or two places.
We imagine the following nice
scenario where 
\begin{eqnarray}
\cale^Q &=& (E_n^\vee, \ldots, E_1^\vee) \nonumber \\
&=&
(\delta F_n[2], \ldots, \delta F_b[2], \delta F_{b-1}[1], \ldots,
\delta F_c[1],\delta F_{c-1}, \ldots, \delta F_1)
\label{shifts}
\end{eqnarray}
and the $F_i$ are sheaves.  Clearly the $\IP^2$
example satisfies this condition.  Indeed all three
block exceptional collections in \cite{NK} also
satisfy this condition, and examples are known 
for all del Pezzos.
We assume the 
preceding blocking structure for $\cale^Q$ and
also that  
$\Hom^k(E_i^\vee, E_j^\vee) = 0$ unless $k=1$ or 2.
We will call these two conditions together 
the $\Ext^{1,2}$
condition.
In the preceding sections, we spent a substantial
amount of time explaining the second condition using
physics considerations.  
%Recall that
%for objects $A$, $B \in D^\flat(\calb)$, 
%\be
%\Hom^r(A[p],B[q]) = \Hom^{r-p+q}(A,B) \ .
%\ee

\begin{lemma}\label{ext12}
The $\Ext^{1,2}$ condition implies well split.
\end{lemma}
\begin{proof}
Consider two sheaves $F_i$ and $F_j$ in 
$\cale^Q$ with the same grading, $i>j$.
Either $\Hom(F_i,F_j) = 0$ or
$\Ext^1(F_i, F_j) = 0$ because
$(F_i, F_j)$ form an exceptional pair.
However, to satisfy the $\Ext^{1,2}$
condition, we must have
$\Hom(F_i, F_j) = 0$.  Thus, the arrow in 
the quiver points from $i$ to $j$.
We conclude there can be no violations
of the well split condition within a block
of sheaves of the same grading.

We can conclude slightly more from the preceding.
In particular, recall that $\chi_-(F_i, F_j) = \chi(F_i, F_j)$.
Thus, the slopes must satisfy $\mu(F_i) \geq \mu(F_j)$.

Next consider $\delta F_i[2]$ and $\delta F_j$ for
$n \geq i \geq b$ and $c-1 \geq j \geq 1$.
By the $\Ext^{1,2}$ condition, 
$\Ext^1(F_i, F_j) = 0$ or we would have an 
$\Hom^3$ type map in $D^\flat(\calb)$.  Thus,
the arrow, if it exists, points from $j$ to $i$.

The only possible violation to well split will occur for
maps between sheaves with gradings shifted by one.
Consider $\delta F_i[2]$ for $n \geq i \geq b$ and
$\delta F_j[1]$, $\delta F_k[1]$ for
$b-1 \geq j > k \geq c$.
Assume $\Ext^1(F_i, F_j) \neq 0$, i.e.
there is an arrow from $j$ to $i$.  We would have
a violation of well split if
then $\Hom(F_i, F_k) \neq 0$, but such a violation
cannot occur for the following reason.
$\Ext^1(F_i, F_j) \neq 0$ implies that
$\mu(F_j) < \mu(F_i)$.  We have from the first
part of the proof that $\mu(F_j) \geq \mu(F_k)$.
Therefore $\mu(F_k) < \mu(F_i)$ and
the map from $F_i$ to $F_k$ must also be
of type $\Ext^1$.   A similar argument rules
out violations of the well split condition for
maps from $F_i[1]$ to $F_j$ in $\cale^Q$.
\end{proof}

We also have a useful corollary for the values of the 
slopes.
\begin{corollary}\label{Qslopes}
For a collection $\cale^Q$ satisfying the $\Ext^{1,2}$
condition, the slopes of the sheaves with a given
grading are in descending order within $\cale^Q$.  
In particular,
for $n \geq i > j \geq b$
or for $b-1 \geq i > j \geq c$ or for
$c-1 \geq i > j \geq 1$, $\mu(F_i)\geq  \mu(F_j)$.
Moreover, $\mu(F_n) \leq \mu(F_1)$.
\end{corollary}

Given an $\cale^Q$, we are assuming that $\cale^Q$ was constructed from
a collection $\cale$ where all the sheaves in $\cale$ are
torsion free.
Otherwise, the corresponding gauge groups would have rank zero,
and we would 
lose our physical motivation for studying
these collections.  It may be that $\cale^Q$ is also
torsion free, but we have not been able to prove it, and it
is therefore worthwhile to consider the special
cases where some of the $E_i^\vee$ have zero rank.

Our proof of Lemma \ref{ext12} holds whether or not
some of the $E_i^\vee$ are torsion.  If $F_i$ were
torsion, we take $\mu(F_i) = \infty$ and $\mu(F_i) \geq 
\mu(F_j)$ $\forall j$.  

Interestingly, the torsion sheaves can only occur in special
places inside $\cale^Q$.
\begin{remark} \label{specialplace}
If $F_i$ is torsion then $i=b-1$ or $i=c-1$ or $F_{i+1}$ is torsion.
\end{remark}
Because $F_n = E_n \otimes K$ is not torsion, the torsion sheaves
will only appear in (\ref{shifts}) shifted by zero or one.

The $\Ext^{1,2}$ property strongly constrains the form
of the original collection $\cale$.  To see these constraints,
we need the following lemma.
\begin{lemma} \label{chernreg}
The chern character of $E_j^\vee$ is
\be
\ch(E_j^\vee) = \sum_{i=1}^j S_{ij} \ch(E_i) \ .
\ee
\end{lemma}
\begin{proof}
From the definition of $S$,
\be
\chi(E_j^\vee, E_i^\vee) = S_{ij} \ .
\ee
From (\ref{eulerdual}),
\be
\chi(E_i, E_j^\vee) = \delta_{ij} \ .
\ee
Because $\cale$ is complete, we can express the chern character of any 
sheaf as a sum over the chern characters of the $E_i \in \cale$.
In particular
\be
\ch(E_j^\vee) = \sum_i a_{ij} \ch(E_i) \ .
\ee
The lemma follows from acting on each side of the equality with
$\chi(\cdot, E_k^\vee)$.
\end{proof}
\begin{corollary}\label{cherninv}
Since $S$ has an inverse, we find
\be
\ch(E_j) = \sum_{i=1}^j S_{ij}^{-1} \ch(E_i^\vee) \ .
\ee
\end{corollary}

\begin{proposition} \label{ext12impse}
If $\cale^Q$ is $\Ext^{1,2}$, then the original collection
$\cale$ generates a strong helix.
\end{proposition}
\begin{proof}
Using Corollary \ref{cherninv}, we have
\begin{eqnarray*}
\mu(E_{j-1}) & \leq & \mu(E_j) \iff \\
\frac{\sum_{i=1}^{j-1} \deg(E_i^\vee) S^{-1}_{i(j-1)} }
{\sum_{i=1}^{j-1} r(E_i^\vee) S^{-1}_{i(j-1)} }
&\leq &
\frac{\sum_{i=1}^{j} \deg(E_i^\vee) S^{-1}_{ij} }
{\sum_{i=1}^{j} r(E_i^\vee) S^{-1}_{ij} } \ .
\end{eqnarray*}
Note that by assumption, all the ranks of the sheaves in
$\cale$ are positive.  We can clear the denominators
without flipping the sign of the inequality.
Thus
\begin{eqnarray*}
\mu(E_{j-1}) & \leq & \mu(E_j) \iff \\
\sum_{i=1}^{j-1} \sum_{k=1}^j S_{i(j-1)}^{-1} S_{kj}^{-1}
( \deg(E_i^\vee) r(E_k^\vee) - \deg(E_k^\vee) r(E_i^\vee)) &\leq & 0 \iff \\
\sum_{i=1}^{j-1} \sum_{k=1}^j S_{i(j-1)}^{-1} S_{kj}^{-1} (S_{ik} - S_{ki}) 
&\leq & 0 \iff \\
-S_{(j-1)j}^{-1} &\leq & 0 \ .
\end{eqnarray*}
The fact that $S$ is upper triangular and has ones on the diagonal
implies that $S_{(j-1)j}^{-1} = - S_{(j-1)j}$.  Moreover, well split tells 
us that $S_{(j-1)j} \leq 0$. 

From Corollary \ref{Qslopes}, we have that $\mu(F_n) \leq \mu(F_1)$.
Note that $F_n = E_n \otimes K$ and $F_1 = E_1$.  Thus,
$\mu(E_n) - K^2 \leq \mu(E_1)$ and the helix must be strongly
exceptional.
\end{proof}

\begin{proposition}\label{seimpext12}
If $\cale$ generates a strong helix, then $\cale^Q$ is
$\Ext^{1,2}$.
\end{proposition}

We will spend the rest of this section in a rather technical proof of this important proposition.
The strategy is first to show that $\cale^Q$ respects the blocking structure of grading 
shifts in 
the definition of $\Ext^{1,2}$.  Next, we will demonstrate that the slopes satisfy
Corollary \ref{Qslopes}.

\begin{claim} By the braid group relations, 
\be
\delta E_j = R_{E_1^\vee}^D \cdots R_{E_{j-1}^\vee}^D E_j^\vee \ .
\ee
\end{claim}
Inverting this relation, we find
\be
E_j^\vee = L_{E_{j-1}^\vee}^D \cdots L_{E_1^\vee}^D \delta E_j  \ .
\ee
The chern character of $E_j^\vee$ can then be expanded as
\begin{eqnarray*}
\ch(E_j^\vee) &=& \ch(E_j) - \sum_{i=1}^{j-1} 
\chi(E_i^\vee, L_{E_{i-1}^\vee}^D \cdots L_{E_1^\vee}^D E_j) \ch(E_i^\vee) \\
&=& \ch(E_j) - \sum_{i=1}^{j-1} \chi(E_i, E_j) \ch(E_i^\vee) \\
&=& \ch(E_j) - \sum_{i=1}^{j-1} S_{ij}^{-1} \ch(E_i^\vee) \ ,
\end{eqnarray*}
which agrees with the result of Corollary \ref{cherninv}.
We have rederived this result to see that each term in the sum corresponds
to a separate left mutation and thus a separate opportunity for the complex $E_j^\vee$
to be shifted by one with respect to $\delta F_j$.
Note that
\be
r(E_j^\vee) = r(E_j) \left[ 1 - \sum_{i=1}^{j-1} r(E_i) r(E_i^\vee) (\mu_j - \mu_i) \right] \ .
\ee
As we sum from $i=1$ to $j-1$, each time the term in brackets changes sign, 
$E_j^\vee$ will shift by one with respect to $\delta F_j$.  Note that
if a partial sum is zero, there is no corresponding shift in grading
because for left mutations a torsion sheaf can only be produced by a recoil,
and there is no shift in grading for a recoil.

Let $c$ be the smallest integer such that $r(E_c^\vee) < 0$.
  There must exist
such a $c$ because $E_n^\vee$ is shifted by two with respect to $\delta F_n$.
For all $i<c$, there is no relative shift and $E_i^\vee = \delta F_i$.  We now
argue that for all $i>c$, $E_i^\vee = \delta F_i [s_i]$ where $s_i$ is either one or two.
The reason is that $\cale$ is strongly exceptional and $\mu_i \leq \mu_{i+1}$.  In particular,
\be
\sum_{i=1}^{c-1} r(E_i) r(E_i^\vee) (\mu_c - \mu_i)  
\leq \sum_{i=1}^{c-1} r(E_i) r(E_i^\vee) (\mu_k - \mu_i) 
\ee
for $k>c$.  Thus, $E_k^\vee$ will be shifted by at least one with respect to
$\delta F_k$.
To summarize, we have found that 
\be
\cale^Q = (\delta F_n[s_n], \ldots, \delta F_{c+1}[s_{c+1}], \delta F_c [1], \delta F_{c-1}, 
\ldots, \delta F_1) \ .
\label{shifts1}
\ee
where $s_i = 1$ or 2.

We now take advantage of the helix structure to complete a proof of the fact
that $\cale^Q$ contains objects with the appropriate shifts of grading.
Using  
\be
\delta E_i \otimes K[2] = L_{\delta E_{i+1}\otimes K[2]}^D \cdots L_{\delta E_n \otimes K[2]}^D
L_{\delta E_1}^D \cdots L_{\delta E_{i-1}}^D \delta E_i \ ,
\ee
we find that
\be
E_i^\vee = R_{\delta E_n \otimes K[2]}^D \cdots R_{\delta E_{i+1}\otimes K[2]}^D 
(\delta E_i \otimes K[2]) \ .
\ee
With the braid relations, 
\be
\delta E_i \otimes K[2] = L_{E_n^\vee}^D \cdots L_{E_{i+1}^\vee}^D E_i^\vee \ ,
\ee
and thus
\be
E_i^\vee = R_{E_{i+1}^\vee}^D \cdots R_{E_n^\vee}^D (\delta E_i \otimes K[2]) \ .
\ee

Just as done above, we can use this expression for $E_i^\vee$ to derive a
 relation on $\ch(E_i^\vee)$:
\be
\ch(E_j^\vee) = \ch(E_j \otimes K) - \sum_{i=j+1}^n S_{ij}^{-1} \ch(E_j^\vee) \ .
\ee
For the ranks, we find then
\be
r(E_j^\vee) = r(E_j) \left[1-\sum_{i=j+1}^n r(E_i) r(E_i^\vee) (\mu_i - \mu_j)
\right]
\ee
which could have been derived from (\ref{cherninv}) and chiral anomaly 
cancellation  ($(S-S^T)\cdot r = 0$), but now again we have an interpretation
of each term in the sum in brackets as a right mutation.

As before, we find that there is a $b$ such that $r(E_i^\vee) > 0$ for
$n\geq i \geq b$ and $r(E_{b-1}^\vee) \leq 0$.\footnote{
For right mutations, a torsion sheaf can be produced by a division and
hence $E^\vee_{b-1}$ may be torsion.}
  Moreover, for
$i<b$, $E_i^\vee = \delta F_i[s_i]$ and $s_i$ is either zero or one.
In particular
\be
\cale^Q = (\delta F_n[2], \ldots, \delta F_b[2], 
\delta F_{b-1}[1],
\delta F_{b-2}[s_{b-2}] \ldots
\delta F_1[s_1]) \ .
\label{shifts2}
\ee
Putting (\ref{shifts1}) and (\ref{shifts2}) together, 
we conclude that the shifts of grading of $\cale^Q$ are precisely as
in (\ref{shifts}).

The last remaining step in the proof of Theorem \ref{sewellsplit} is to 
check that the slopes are as in 
Corollary \ref{Qslopes} (which in turn implies the restriction on
the types of $\Ext$ maps between the sheaves).
We first consider the $E_i^\vee$ which are torsion free. 
Notice that $\sign(r(E_i^\vee)) = \sign(r(E_j^\vee))$ for
$n\geq i > j \geq b$ or for $b-1 \geq i > j \geq c$ or
for $c-1 \geq i > j \geq 1$.
Using Lemma \ref{chernreg}, we have
\begin{eqnarray*}
\mu(E^\vee_{j-1}) & \leq & \mu(E^\vee_j) \iff \\
\frac{\sum_{i=1}^{j-1} \deg(E_i) S_{i(j-1)} }
{\sum_{i=1}^{j-1} r(E_i) S_{i(j-1)} }
&\leq &
\frac{\sum_{i=1}^{j} \deg(E_i) S_{ij} }
{\sum_{i=1}^{j} r(E_i) S_{ij} } \ .
\end{eqnarray*}
We can clear the denominators without flipping the sign of the inequality
only if $\sign(r(E_{j-1}^\vee)) = \sign(r(E_j^\vee))$, which will be true if $E_j^\vee$
and $E_{j-1}^\vee$ are shifted by the same amount.
Manipulations analogous to those in the proof of Proposition
\ref{ext12impse} show 
\begin{eqnarray*}
\mu(E^\vee_{j-1}) & \leq & \mu(E^\vee_j) \iff \\
\sum_{i=1}^{j-1} \sum_{k=1}^j S_{i(j-1)} S_{kj}
( \deg(E_i) r(E_k) - \deg(E_k) r(E_i)) &\leq & 0 \iff \\
\sum_{i=1}^{j-1} \sum_{k=1}^j S_{i(j-1)} S_{kj} (S_{ki}^{-1} - S_{ik}^{-1}) 
&\leq & 0 \iff \\
S_{(j-1)j} &\leq & 0 \ .
\end{eqnarray*}
From the structure of these upper triangular matrices, 
$S_{(j-1)j} = -S_{(j-1)j}^{-1}$.  From the
strong exceptional property, $S_{(j-1)j}^{-1} \geq 0$.

We also check that $\mu(F_n) \leq \mu(F_1)$.
Note that $\mu(F_n) = \mu(E_n \otimes K) = \mu(E_n) -K^2$
and $\mu(F_1) = \mu(E_1)$.
But we know that $\mu(E_n) \leq \mu(E_1) + K^2$ because
 $\cale$ generates a strong helix.

Finally, we consider the $E_i^\vee$ which are torsion.  If the torsion
sheaves occur as in Remark \ref{specialplace}, then $\cale^Q$ continues
to satisfy the conditions of Corollary \ref{Qslopes} because the slope
of a torsion sheaf is infinite.  Note that $F_1 = E_1$ and 
$F_n = E_n \otimes K$ are torsion free from our assumption about $\cale$.
We can then finish our proof of Proposition \ref{seimpext12} by showing
that for $E_i^\vee$ and $E_{i+1}^\vee$ shifted by the same amount,
if $F_{i}$ is torsion, then so is $F_{i+1}$:
\begin{eqnarray*}
S_{i(i+1)} = 
\chi(E_{i+1}^\vee, E_i^\vee) = \chi(F_{i+1}, F_i) = r(F_{i+1}) \geq 0 \ .
\end{eqnarray*}
However, as above we have that $S_{i(i+1)} = -S^{-1}_{i(i+1)} \leq 0$
and we conclude that $r(F_{i+1})=0$.

\subsection{Proof of Theorem \ref{setheorem}}

Let $\cale$ generate a strong helix.
We are free to take the Seiberg dual of $E_n$ without
loss of generality.  (We just choose an appropriate foundation of the helix.
The strong property of the helix means that any foundation will also be strongly
exceptional.)
Let $\cale'$ be the Seiberg dual collection:
\be
\cale' = (E_1, \ldots, E_a, L_{E_{a+1}} \cdots L_{E_{n-1}} E_n, E_{a+1}, \ldots, E_{n-1}) 
\label{eprimedef}
\ee
where we have assumed that nodes 1 through $a$ all have arrows ending on node $n$
while nodes $a+1$ through $n-1$ have arrows beginning at node $n$.
Note that from the helix property,
\be
E_n' \equiv L_{E_{a+1}} \cdots L_{E_{n-1}} E_n = R_{E_1} \cdots R_{E_a} (E_n \otimes K) \ .
\label{alternateE1}
\ee

Our strategy is to
show that the slopes of $\cale'$ satisfy the conditions of Remark \ref{strslopes}.
$\cale'$ is an exceptional collection, and the conditions on the 
slopes are sufficient to guarantee that 
$\cale'$ is strongly exceptional and in fact generates a strong helix.

\begin{lemma}\label{chernone}
The chern character of $E_n'$ is
\be
\ch(E_n') = -\sum_{i=a+1}^n S_{in} \ch(E_i) \ .
\ee
\end{lemma}
\begin{proof}
From the definition of
mutation in terms of short exact sequences, we find that
\be
\ch(E_n') = (-1)^{\delta_{n-1}} \ch(E_n) + 
\sum_{i=a+1}^{n-1} (-1)^{\delta_i} S_{in}' \ch(E_i) \ 
\ee
where we have defined 
\be
S_{in}' = -\chi(E_i, L_{E_{i+1}} \cdots L_{E_{n-1}} E_n) \ 
\ee
and $\delta_i$ is the number of mutations among the $L_{E_{a+1}}, \ldots, L_{E_i}$
of type division.

Recall the definition
\be
S_{jn} = \chi(E_n^\vee, E_j^\vee) = 
(L_{\delta E_1}^D \cdots L_{\delta E_{n-1}}^D \delta E_n, L_{\delta E_1}^D \cdots 
L_{\delta E_{j-1}}^D 
\delta E_j) \ .
\ee
We can massage this formula into something a little more useful
\begin{eqnarray*}
S_{jn} &=& \chi(L_{\delta E_j}^D \cdots L_{\delta E_{n-1}}^D \delta E_n, \delta E_j) \\
&=& -\chi(\delta E_j, L_{\delta E_{j+1}}^D \cdots L^D_{\delta E_{n-1}} \delta E_n) \\
&=& (-1)^{\epsilon_j} S_{jn}'
\end{eqnarray*}
where $\epsilon_i$ is number of mutations of type division among the
$L_{E_{i+1}}, \ldots, L_{E_{n-1}}$.  Clearly, $\epsilon_i+\delta_i$ is a constant.
In particular, $\epsilon_i + \delta_i = \epsilon_{a} = \delta_{n-1}$.
We conclude that
\be
\ch(E_n') = (-1)^{\epsilon_a} \left[ 
\ch(E_n) + \sum_{i=a+1}^{n-1} S_{1i} \ch(E_i) 
\right]
\ .
\ee

We argue that $\epsilon_a = 1$.  
$\epsilon_a = 0$, $1$, or $2$
and, from (\ref{leftshift2}),  there would be exactly two divisions if we were to left mutate
$E_n$ through the whole collection.
For an exceptional pair $(F,E)$, recall that a division occurs
whenever $r(L^D_{\delta F} \delta E) = r(E) - \chi(E,F) r(F)$ is negative.  
The well split condition implies
that $S_{in} \geq 0$ for $1 \leq i \leq a$ and $S_{in} \leq 0$ for
$a+1 \leq i \leq n-1$.  
In order to have two divisions as a result of mutating
$E_n$ through the collection, from the signs of the
$S_{in}$, we need exactly one division by the time
we get to $i=a+1$.  Therefore $\epsilon_a = 1$.
\end{proof}
%We conclude that
%\be
%\ch(E_1') = -\ch(E_1) - \sum_{i=2}^a S_{1i} \ch(E_i) \ .
%\ee

Note that $r(E_n') > 0$ and $E_n'$ cannot be torsion because
the grading does not shift if a left mutation produces a torsion
sheaf.  

It follows from Lemma \ref{chernone} that
the slope of $E_n'$ is
\be
\mu(E_n') = \frac{\deg(E_n')}{r(E_n')} = 
\frac{- \sum_{i=a+1}^{n} S_{in} \deg(E_i)}
{ - \sum_{i=a+1}^{n} S_{in} r(E_i)} \ .
\ee

\begin{lemma}\label{ineqone}
The slope of $E_n'$ satisfies the inequality:
\be
\mu(E_n') \geq \mu(E_a) \ .
\ee
\end{lemma}
\begin{proof}
%We show that $\mu(E_1') \leq \mu(E_{a+1})$:
\begin{eqnarray}
\mu(E_n') &\geq& \mu(E_a) \iff \nonumber \\
\sum_{i=a+1}^n S_{in} \left(
r(E_i) \deg(E_a) - r(E_a) \deg(E_i) 
\right) & \geq& 0 \iff \nonumber \\
-\sum_{i=a+1}^n S_{ai}^{-1} S_{in} &\geq& 0 \iff \nonumber \\
S_{an} &\geq& 0 \ .
\label{ineq}
\end{eqnarray}
By the well split assumption, $S_{an}\geq 0$.
\end{proof}

\begin{lemma}\label{ineqtwo}
The slope of $E_n'$ satisfies the additional
inequality:
\be
\mu(E_n') \leq \mu(E_{a+1}) \ .  
\ee
\end{lemma}
\begin{proof}
It is convenient to choose a different foundation for the helix to
check the inequality.  
%In so doing, we require that the helix and not
%just the collection is strongly exceptional.
We rename
$E_n \otimes K = E_0$, and then we introduce
\be
\calf = (F_1, \ldots, F_n) = (E_0, E_1, \ldots, E_{n-1}) \ .
\ee
Using (\ref{alternateE1}) and an argument
analogous to the proof of
Lemma \ref{chernone},
we find that
\be
\ch(E_n') = - \sum_{i=1}^{a+1} S_{1i} \ch(F_i) \ 
\ee
where now the $S_{ij}$ have been defined with respect
to $\calf$.
It follows that
\be
\mu(E_n') = \frac{- \sum_{i=1}^{a+1} S_{1i} \deg(F_i) }
{- \sum_{i=1}^{a+1} S_{1i} r(F_i)} \ .
\ee
One additional subtlety is the fact that we are using right mutations
to construct $E_n'$.  Thus we need the fact from the proof
of Lemma \ref{chernone} that $r(E_n') \neq 0$.
An argument equivalent to (\ref{ineq}) demonstrates
that $\mu(E_n') \leq \mu(F_{a+2}) = \mu(E_{a+1})$
if and only if $S_{1(a+2)} \geq 0$, which follows from 
well split.
\end{proof}

From Lemmas
\ref{ineqone} and \ref{ineqtwo}, it
follows that $\cale'$ is strongly exceptional.

Now we check that $\cale'$ generates a strong
helix.
Note that $\mu(E_n) - K^2 \leq \mu(E_1) \leq \mu(E_2)$
so $\mu(E_n) - K^2 \leq \mu(E_2)$.

\section{Recovering the Old Seiberg Duality}
\label{sec:usualdef}

As promised, we discuss how our definition of Seiberg duality matches
the original definition.

For the original Seiberg duality, one takes a node $i$ of a supersymmetric quiver
and reverses the orientation of all the arrows incident on that node, $X_{ij} \to
X_{ji}'$ and $Y_{ji} \to Y_{ij}'$.  The old maps $X_{ij}$ and $Y_{ki}$ one combines
to make so called mesonic fields (or maps) $M_{kj} = X_{ij} Y_{ki}$.  To the superpotential,
one adds new mass terms $W \to W + M_{kj} Y_{ik}' X_{ji}' $ for each $M_{kj}$ field.

Sometimes, the $M_{kj}$ maps will be oriented opposite to the maps $X_{kj}$ or $X_{jk}$
present
in the original quiver.  If we hope to be able to represent the quiver with
exceptional collections, we need to be able to ensure that the maps between $k$ and
$j$ are only in one direction.  One can ``integrate out" the $M_{kj}$
(or equivalently the $X_{jk}$),
by which is meant one uses the relations $\partial W / \partial M_{kj} = 0$ 
to eliminate $M_{kj}$ from W.  Such a procedure is not always possible.  

Let's compare this procedure with exceptional collections.
We Seiberg dualize on node 1, obtaining
the collection $\cale'$ given by
\be \label{neweprimedef}
\cale' = (E_2, \ldots, E_a, R_{E_a} \cdots R_{E_2} E_1, E_{a+1}, \ldots, E_n) 
\ee
 
\begin{proposition}\label{dualprime}
The dual exceptional collection ${\cale^Q}'$ takes the form
\be
{\cale^Q}' = 
(E_n^\vee, \ldots, E_{a+1}^\vee, \delta E_1[1], R_{\delta E_1}^D E_a^\vee,
\ldots, R_{\delta E_1}^D E_2^\vee) \ .
\ee
\end{proposition}
\begin{proof}
The result follows from the expression for $\cale'$ (\ref{neweprimedef}),
the definition of $\cale^Q$ (\ref{dualdef}), and the braid group relations
(\ref{braidrels}). 
\end{proof}
From this Proposition, we see that the complex $\delta E_1$ 
corresponding to the dualized node gets shifted by one under Seiberg duality,
reversing the orientation of all the arrows incident on node $1$.  Moreover,
the number of arrows from $k$ to $j$ is shifted by
\be
S_{kj} \to \chi(E_j^\vee, R^D_{\delta E_1} E_k^\vee) =
S_{kj} - S_{1k} S_{1j}
\ee
If $S_{kj} > 0$, the number of arrows between $k$ and $j$ matches the
original definition of Seiberg duality.  If $S_{kj}<0$, we have to assume
in the original picture of Seiberg duality that 
we can integrate out enough of the $M_{kj}$ or $X_{jk}$ such that
the result has maps only in one direction.

The original definition of Seiberg duality also came with an induced action
on the quiver representation, often denoted by physicists as $N_c \to N_f - N_c$
where $N_c$, the number of colors, is the rank $d^i$
of the
vector space at the dualized node $i$ and $N_f$ is the number of flavors.
In particular, if we were to dualize node 1 say,
\be
N_f = -\sum_{i=2}^a S_{1i} d^i \ .
\ee
In other words, we recover
\be
d^1 \to -\sum_{i=1}^a S_{1i} d^i
\ee
in agreement with Lemma \ref{chernone}.

\section{Discussion}

The results in this paper help to resolve a number of bothersome puzzles and raise
some interesting questions.  As we discuss presently, these puzzles concern 1)
the recipe for constructing a gauge theory from an exceptional collection, 2) 
negative conformal dimensions of gauge invariant operators, 3) the
connection between the exceptional collection literature in mathematics
and del Pezzo gauge theories in physics, and 4) the connection between
different mathematical formulations of Seiberg duality.

The original recipe for writing down a gauge theory quiver from an exceptional collection
presented in section 4 
now makes a lot more sense if we assume the collection generates a strong helix.
The original recipe for writing down a gauge theory quiver from an exceptional collection
 relied only on the sign of the Euler character, seeming to ignore
the significance of whether the maps between the sheaves were of type
$\Ext^0$, $\Ext^1$, $\Ext^2$, or $\Ext^3$.  From the discussion in section 2, we expected
that only $\Hom_{D^\flat(\bX)}^1(E,F)$ type maps should correspond to 
bifundamental fields in the gauge theory.  If the collection generates a strong helix, 
we find that there will only be $\Hom_{D^\flat(\bX)}^1$ type maps and the sign
of the Euler character just lets us know whether the map descended from
a $\Hom^1_{D^\flat(\calb)}$ or $\Hom^2_{D^\flat(\calb)}$ in the del Pezzo.

The recipe makes more sense not only in a physics context but also a mathematical one.
We can think about the gauge theory quiver as generating an algebra, the 
path algebra of the quiver.  This path algebra is nothing but the maps between the
sheaves in the helix $\calh$, at least as long as $\calh$ is a strong helix.
However, if there are higher $\Ext$ maps between the sheaves in $\calh$,
it is far less clear to what this path algebra of the quiver corresponds.

Concerns have been raised in the literature \cite{Herzog} 
about possible gauge invariant,
chiral operators with negative R-charge (and hence negative conformal dimension)
which our results here eliminate.
The authors of \cite{HW}, inspired in particular by work of \cite{IW2}
but also by \cite{HM}, give a formula for
the R-charges of the bifundamental fields $X_{ij}$.  If $i<j$, then either
\be
\frac{K^2}{2} R(X_{ij}) = \mu_j - \mu_i
\ee
or
\be
\frac{K^2}{2} R(X_{ji}) = K^2 - \mu_j + \mu_i \ ,
\ee
depending on whether the arrow goes from $i$ to $j$ or from $j$ to $i$.
For general exceptional collections, $R(X_{ij})$ may be less than zero.  While
not of concern by itself, gauge invariant combinations of the these $X_{ij}$,
called dibaryon operators, can be made which also have negative R-charge.
However, for strong helices
\be
0 \leq \mu_j - \mu_i \leq K^2
\ee
and hence the R-charges will always be positive.

Our results demonstrate the physical relevance
of certain mathematical concepts and we hope may 
serve as inspiration to those who study
exceptional collections for their own sake.  
Unaware of the connection to gauge theory,
Bondal \cite{Bondal} defined and studied 
strong helices and admissible mutations.
For example, Corollary 7.3 of \cite{Bondal} 
states that if $\cale^Q$ is strongly
exceptional and Koszul, then $\cale$ is strongly exceptional.
In this paper, we have replaced Corollary 7.3 with an equivalence
between $\Ext^{1,2}$ and strong helices.  
Section 8 of \cite{Bondal} is a preliminary investigation of admissible mutations
of strong helices.  Here, we have been able to argue that
an important class of these admissible mutations are Seiberg dualities.
The precise connection between Koszul and $\Ext^{1,2}$ and whether
or not there are other admissible mutations in addition to Seiberg dualities
deserve further thought.

These admissible mutations of strongly exceptional collections
must be closely related to the tilting equivalences of
Berenstein and Douglas \cite{BD}.
Bondal \cite{Bondal} 
constructs a quiver from strongly exceptional collections.
The quiver he constructs, the Beilinson quiver of section 4,
 is not quite the gauge theory
quiver but is closely related to it.  He is then
able to prove (Theorem 6.2 of 
\cite{Bondal}) that for strongly exceptional collections,
$D^\flat(\calb)$ is equivalent to the derived category
constructed from the quiver.  
We suspect there must be a stronger theorem
which relates the derived category constructed
from the gauge theory quiver to $D^\flat(\bX)$ and that
 in this context tilting equivalences will prove to be precisely
admissible mutations.

There are at least two important physics questions that we have not been able to address here.
One is the existence of an analog of the 
orbifold point in the Kaehler moduli space of $\bX$ where
all the D-branes become mutually supersymmetric.  Without such
a point or locus, our gauge theory construction fails.  We think it likely such
a point exists, but there are no guarantees,  and 
a more careful analysis of the Kaehler moduli space
of these del Pezzos is in order.  
Two is whether anything we have learned here about strongly exceptional collections 
can be applied to the other physical situations where exceptional collections are important,
for example the Landau-Ginzburg models of \cite{zaslow, hiv}.
We leave these questions for the future.

\flushleft{Note added: After the electronic distribution of the first version of this 
work, \cite{Aspinwallnew} appeared which has some overlap with our results.}

\section*{Acknowledgments}

I would like to thank Paul Aspinwall, Brian Conrad, 
Igor Dolgachev, and David Gross, 
for useful conversations.
I would also especially like to thank 
David Berenstein, James M\raise 3pt \hbox{\text {\normalsize c}}Kernan, 
and Johannes Walcher for comments on the manuscript.
Finally, I am grateful to the referee for insisting on a more careful treatment
of torsion sheaves.
This research was supported in 
part by the National Science Foundation under Grant No.
PHY99-07949.

\end{document}